# Small Animal Multivariate Brain Analysis (SAMBA) – A High Throughput Pipeline with a Validation Framework

Robert J. Anderson[1], James J. Cook[1], Natalie Delpratt[1,5], John C. Nouls[1], Bin Gu[2,6], James O. McNamara[2,3,4], Brian B. Avants[7], G. Allan Johnson[1,5], Alexandra Badea*[1, 5]

[1]Center for In Vivo Microscopy, Department of Radiology
[2]Department of Pharmacology and Cancer Biology
[3]Department of Neurobiology
[4]Department of Neurology
Duke University Medical Center, Durham, NC 27710, USA
[5]Department of Biomedical Engineering
Duke University, Durham, North Carolina, USA
[6]Department of Cell Biology and Physiology
University of North Carolina, Chapel Hill, NC 27599, USA
[7]Biogen, Cambridge, MA, 02142, USA

**Correspondence:**

Alexandra Badea, PhD
Box 3302 Duke University Medical Center
Durham, NC, 27710, USA
Tel: +1 919 684-7654
alexandra.badea@duke.edu

9 Figures; 3 Tables

**Abstract**

While many neuroscience questions aim to understand the human brain, much current knowledge has been gained using animal models, which replicate genetic, structural, and connectivity aspects of the human brain. While voxel-based analysis (VBA) of preclinical magnetic resonance images is widely-used, a thorough examination of the statistical robustness, stability, and error rates is hindered by high computational demands of processing large arrays, and the many parameters involved. Thus, workflows are often based on intuition or experience, while preclinical validation studies remain scarce. To increase throughput and reproducibility of quantitative small animal brain studies, we have developed a publicly shared, high throughput VBA pipeline in a high-performance computing environment, called SAMBA. The increased computational efficiency allowed large multidimensional arrays to be processed in 1-3 days—a task that previously took ~1 month. To quantify the variability and reliability of preclinical VBA in rodent models, we propose a validation framework consisting of morphological phantoms, and four metrics. This addresses several sources that impact VBA results, including registration and template construction strategies. We have used this framework to inform the VBA workflow parameters in a VBA study for a mouse model of epilepsy. We also present initial efforts towards standardizing small animal neuroimaging data in a similar fashion with human neuroimaging. We conclude that verifying the accuracy of VBA merits attention, and should be the focus of a broader effort within the community. The proposed framework promotes consistent quality assurance of VBA in preclinical neuroimaging; facilitating the creation and communication of robust results.

## 1. Introduction

Computational imaging has emerged as a powerful neuroscience research tool. It has been used to identify patterns of human brain differences due to genotype, environment (Blokland et al., 2012), development (Becker et al., 2016), aging (Kremen et al., 2013), and disease (Thompson et al., 2014). The reliability of such analyses has received increased attention and scrutiny in human brain neuroimaging (Shen and Sterr, 2013) (Radua et al., 2014; Michael et al., 2016) (Eklund et al., 2016). Exploring such themes in rodents provides important insight into human conditions, as phenotypes can be replicated via genetic manipulation, while environmental and other conditions can be well controlled. Indeed, murine models of neurologic diseases have played a critical role in neuroscience. It is thus crucial to develop accurate and reliable techniques specific to small animal imaging.

Our main objective is neuroanatomical phenotyping using MR histology (Johnson et al., 1993). Diffusion tensor imaging (DTI) is an attractive tool for MR histology, as it delivers multiple contrasts such as fractional anisotropy (FA) and radial diffusivity (RD) to quantify microstructural integrity (Calabrese et al., 2014). Additionally, using DTI contrasts to drive image registration can improve the resulting alignment (Badea et al., 2012). We thus need the ability to handle multiple contrasts.

Voxel-based analysis (VBA) has been established as a method for localizing and quantifying morphometric and physiological brain changes (Ashburner and Friston, 2000). VBA has been used with magnetic resonance imaging (MRI), positron-emission tomography (PET), and single-photon emission computed tomography (SPECT) (Good et al., 2001; Hayasaka et al., 2004). Among these, MRI is particularly well-suited for anatomical phenotyping in small animals (Johnson et al., 2002; Nieman et al., 2005; Badea et al., 2007; Johnson et al., 2007; Borg and Chereul, 2008; Badea et al., 2009; Ellegood et al., 2015). MRI based VBA in mice has provided unique insights into conditions such as Huntington's (Sawiak et al., 2009b), Alzheimer's Disease (Johnson et al., 2010), or the effects of exercise (Biedermann et al., 2012); and the number of VBA studies VBA continues to grow.

A major issue with VBA is the long computational time. In its most critical step, spatial normalization is realized by registering each subject to a common template. Diffeomorphic Symmetric Normalization (SyN) (Avants et al., 2008) has become the algorithm of choice for this task since Klein et al. (2009) found that it outperforms other approaches – in people. A typical clinical exam features T1- , T2- or T2*-weighted scans with 1-mm isotropic voxel size, and 256x256x200 arrays, yielding about 25 MB per/scan or 75 MB per set. A DTI scan in ADNI uses a 128x128x59 array, 41 diffusion directions and 5 non diffusion weighted scans, and will produce 85MB (ADNI, accessed 5/30/2017) . In contrast, rodent brain MRI acquisitions are substantially larger (Johnson et al., 2012; Lerch et al., 2012; Calabrese et al., 2015b), and may include gradient-recalled echo (GRE) sequences at 21 μm isotropic resolution, using 1024x512x512 arrays (512 MB); and DTI protocols at 43 μm resolution, using 512x256x256 arrays. The resulting DTI parametric images, e.g. FA, are 8.5 times larger for one mouse brain relative to the human; and sum up to ~ 1 GB per specimen for all 7 of the standard DTI contrasts. Our multivariate analysis VBA pipeline thus needs to handle ~15 times more data than the 66 MB required for structural voxel based morphometry in humans based in T1/T2 protocols. For such large arrays high-quality SyN registrations come with higher price tags, as a single registration of two mouse brains at 56 μm isotropic resolution can take ~100 CPU hours (VanEede et al., 2013). The best-case scenario, from a processing time perspective, would be to select one subject as the target template, requiring only ($N$-1) registrations. But this introduces a bias towards the selected specimen. To eliminate bias a better practice is to construct a study-specific, minimum deformation template (MDT) (Kochunov et al., 2001; Avants et al., 2010). Even an efficient iterative MDT strategy requires at least 3 iterations of pair-wise registrations between each MDT-contributing subject and the target template, a minimum of 3*$N_{MDT}$ jobs. Then all subjects need to be registered to the final MDT, for a total of 4*$N_{MDT}$ jobs. Consider a relatively small study consisting of 10 control ($N_C$) and 10 treated ($N_T$) mice, where only the controls are used to create the MDT. A total of 3*$N_C$+($N_C$+$N_T$) = 50 jobs are necessary, or ~ 30 weeks of CPU time in the hypothetical scenario that a single CPU would be used. The numbers become more daunting as the number of subjects increases. It is therefore imperative to identify and implement efficient computational strategies for MRI-VBA.

Among possible solutions, single workstations are limited in processing power and memory. While cloud computing presents an attractive strategy, significant effort is required upfront to set up processing pipelines. Computing time, data transfer and storage are all issues to be addressed. Here we present a local cluster implementation of an automated processing that adopts parallel processing of locally stored data.

An automated processing pipeline should ensure a reproducible, tractable workflow. It also saves time by reducing human interaction, which can introduce errors, especially when many processing steps are involved. Multiple pipelines have been designed for human brain imaging, including: the FMRIB Software Library (FSL) (Smith et al., 2004; Jenkinson et al., 2012), Statistical Parameter Mapping (SPM) (Friston et al., 1994), and the LONI pipeline (Dinov et al., 2009). These however, do not translate immediately to the preclinical domain, due to difference in scale, gray/white matter distributions and contrasts, and a lissencephalic rodent brain. Image-processing pipelines for preclinical MRI neuroanatomical phenotyping have also been developed for automatic registration (Friedel et al., 2014), segmentation (Johnson et al., 2007; Badea et al., 2009; Minervini et al., 2012), label-based analysis (LBA) (Borg and Chereul, 2008; Budin et al., 2013), cortical thickness (Lerch et al., 2008; Lee et al., 2011), and VBA/voxel-based morphometry (Sawiak et al., 2009b; Lerch et al., 2011; Sawiak et al., 2013; Calabrese et al., 2014). Recently Pagani et al. (2016) described a pipeline that integrates all these four functions. Little attention has however been given to evaluating computational costs, which can be drastically reduced by a high-performance computing (HPC) implementation. Given the increased array sizes, it is essential to have access to sufficient hard drive and memory (RAM) resources, which even high-end workstations may not deliver. Computing clusters provide opportunities for increasing throughput for large numbers of independent tasks (Dinov et al., 2010; Frisoni et al., 2011), as is the case for VBA. Thus, we here propose a high thruput processing pipeline for small animal multivariate brain analysis: SAMBA. SAMBA takes advantage of high-performance computing (HPC), and is based on the widely used Advanced Normalization Tools package (ANTs) (Avants et al., 2009; Avants et al., 2014).

HPC clusters can handle massive amounts of parallel image registrations. VanEede et al. (2013) elegantly demonstrated this by completing 14½ *years*' worth of CPU processing in approximately 2 months. However, the strength of a processing pipeline lies not just in speed. This speed enables us to produce reliable and repeatable results, and to address an unmet need for validation. Verifying VBA accuracy in preclinical studies is paramount, given the increased computational demands. Because VBA comprises multiple processing stages (e.g. spatial normalization, smoothing, and statistical analysis), even small differences in the analysis can lead to divergent conclusions (Rajagopalan and Pioro, 2015). Notably, Bookstein (2001) pointed out that besides physiological sources, statistically significant effects can also arise due to missregistration. In addition, there are ongoing debates about the methodology (e.g. for registration, modulation, and statistical analysis) Thacker (2005). These concerns can be partially addressed by visual inspection, focusing on segmented structures, or cross-validating with other modalities. Here we address the need for a quantitative substantiation of VBA studies and propose a formal validation framework.

The VBA pipeline is most sensitive to changes in the processing chain in voxel based morphometry (VBM). Here, voxel-wise volumetric differences are calculated from the determinant of the Jacobian matrix of the deformation fields mapping each individual to a target template (Chung et al., 2001). This contrast directly encodes local volumetric information, after compensating for global changes. Access to a VBM ground truth "phantom" will enhance any quantitative validation of the system-wide performance. However, no gold standard for preclinical VBM exists. In the clinical domain (Camara et al., 2006; Karaçali and Davatzikos, 2006) have simulated atrophy or hypertrophy to explore how registration affects the sensitivity of deformation recovery, an approach adapted by (VanEede et al., 2013) for the mouse brain as well. Here we propose a set of phantom images, which can be tuned to the expected deformation in a study, used to guide the selection of pipeline parameters, and to estimate the accuracy of the VBA results.

We show the example of phenotyping a mouse model of temporal lobe epilepsy (Lévesque and Avoli, 2013). In this model kainic acid (KA) is injected in the right basolateral amygdala, resulting in epileptic seizures, hippocampal neurodegeneration, and gliosis (Ben-Ari et al., 1980; Mouri et al., 2008; Liu et al., 2013), granule cell dispersion, and mossy fiber sprouting. Accurately recovering these changes presents a non-trivial challenge. We illustrate how VBA/VBM results span a surprisingly wide range, when varying the pipeline parameters. These choices can be informed by phantom metrics, and underscore once again the need for validation.

To address the need for valid statistical analyses, shown clearly for human fMRI (Eklund et al., 2016; Jovicich et al., 2016), but also morphometry (Hosseini et al., 2016) we incorporate several tools for parametric and nonparametric analysis in our pipeline. This, together with the automated documenting of the processing chain allow for further optimization and validation studies, and encourage best practices adoption (Nichols et al., 2017) for small animal imaging.

Our contributions include: 1) the development of a cluster-based VBA pipeline for multi-modal preclinical imaging; 2) an evaluation of the time efficiency gained from parallelizing the pipeline tasks; 3) a validation framework consisting of morphological phantoms and VBA-specific metrics; 4) an examination of how phantom studies inform the parameter selection, and exemplify the consequences of such selections using a mouse model of temporal lobe epilepsy. These datasets are organized in a manner that parallels recent human neuroimaging standardization efforts (Gorgolewski et al., 2016). While parallel computing is a common strategy in image processing, its adoption to VBA in small animal imaging has been limited. As the HPC implementation led to significant gains in processing efficiency, it also enabled a thorough exploration of multiple parameter sets, and evaluation metrics; both in synthetic phantoms, as well as in the case of a mouse model of epilepsy

## 2. Materials and Methods

*2.1. Software, hardware, and pipeline overview*

The VBA pipeline is scripted in Perl and built with a flexible modular structure. In addition to Advanced Normalization Tools (ANTs), software called by the pipeline include MATLAB® (The MathWorks, Inc., Natick, MA), SurfStat (Worsley et al., 2009), FSL, and the R statistical programming language (R Core Team, 2015) with the Advanced Normalization Tools for R (ANTsR) package (Avants et al., 2015). Unless otherwise specified, all commands mentioned herein are from the ANTs toolkit (version/commit date: 13 October 2014 https://sourceforge.net/projects/advants/), and the *antsRegistration* command is used for all registration jobs. The pipeline runs on a Dell HPC cluster featuring the RedHat Enterprise Linux 6.7 operating system, managed via Bright Cluster Manager with Simple Linux Utility for Resource Management (SLURM) (Yoo et al., 2003) for job scheduling and resource allocation. The cluster consists of 11 nodes: a master node, and 10 CPU children nodes (Intel Xenon E5-2697), one of which offers GPU capabilities. Each child node features 16 logical cores (32 via hyper-threading) and 256-GB RAM, with a 4.2-TB hard drive system spread in redundancy across pairs of nodes, yet with the data equally accessible to all nodes.

We highlight here the key elements of the pipeline, while additional discussions of the VBA processing can be found in Supplementary Material. **Figure 1** outlines the main stages of the pipeline, which handles multi-modal data, which may or may not be co-registered, e.g. MR-DTI (upper-left inset), or CT with PET respectively. Additional input required from the user is entered via a matrix of predictors, and a headfile. The matrix of predictors contains metadata of the study's subjects (e.g. specimen ID, MR run number(s), age, gender, treatment; while the headfile is a text file including relevant processing parameters and variables. An extensive input data check is performed, and default values are assigned to missing parameters.

Stage 1 ensures the consistency and quality of the data before launching long-running jobs. Spatial consistency is achieved through recentering, setting the desired orientation, and enforcing a common reference space for all images. All intra-subject acquired contrasts are rigidly co-registered, using the Mutual Information (MI) similarity metric. Quality-related tasks include bias field correction (if appropriate) and skull-stripping (Badea et al., 2007).

Inter-subject linear alignment is performed during Stages 2 and 3. All images are first rigidly aligned to an atlas, or standard coordinate system, such as Waxholm Space ("WHS") (Johnson et al., 2010). After that, they are affinely aligned to a study-based target image. This can be one of the controls, or an unbiased average linear template (ALT) created from all study subjects. Constructing an ALT is based on pairwise registration with an intermediate template (B). The Mattes similarity metric is used for all linear registrations (Mattes et al., 2003).

Stages 4, 5, and 6 are the computational bottlenecks of the pipeline (D). Here, diffeomorphic SyN registrations, which have a large number of parameters, are used to non-linearly spatially normalize all subjects to an MDT.  The strategy for creating the MDT optimizes both shape and appearance (Avants et al., 2010), and is based on the *antsMultivariateTemplateConstruction2* script. This approach starts with an affine average of the images as the initial target template. In each iteration, all subjects selected to construct the MDT (the "MDT group") are registered to the template from the previous iteration (Stage 4). Once all pairwise registration jobs are complete, each "to-MDT" warp—the forward warp when the MDT is considered "fixed"—is applied to their respective images, which are subsequently averaged to produce an intermediate template optimized by appearance. To optimize by shape, all "to-MDT" warps are averaged and multiplied by a small negative constant.  The result is treated as a velocity field which is composed $k$ times (typically 4) to produce a first-order estimate of the inverse average diffeomorphism.  When this transformation is applied to the current template, it has the effect of diffeomorphically moving its shape closer to a location requiring the minimal amount of deformation across all of the MDT group (making it a true "MDT"). This is the final product of the MDT creation iteration. The MDT approaches a stable state in 3-10 iterations (Avants et al., 2010). Here we used 6 iterations, the first 3 of which are decreasingly down-sampled (i.e. *i1* is performed at 8x downsampling, *i2* at 4x, *i3* at 2x, and *i4–i6* at full sampling). This dramatically reduces computation time, while improving the template estimate at each successive resolution. Once the final template has been created, all subjects are independently re-registered to it in Stage 5. This minimizes bias towards the subjects in the MDT group. For non-linear registrations between like contrasts the cross-correlation (CC) similarity metric is used, while mutual information (MI) is used for unlike contrasts. Finally, the diffeomorphic warps from Stage 5, and the previously calculated rigid and affine matrices are used to map the original images into the MDT space.

Stage 6, Label-Based Analysis (LBA), consists of atlas-based segmentation (Gee et al., 1993). Label sets are generated via affine and diffeomorphic registration between the MDT and a labeled brain. The atlas label set is propagated to the MDT, and then to all individuals with the *antsApplyTransforms* command. A MATLAB script is used to calculate for each label its mean volume and mean values of the various contrast intensities, for each subject. Study-wide regional statistics are then computed in conjunction with the matrix of predictors.

In Stage 7 a mask derived from the MDT is eroded with a kernel of the same size as the largest smoothing kernel used, in order to avoid spurious voxels near the mask boundary. We used a 3 voxel kernel, corresponding to ~150 microns, which has worked well in our experience but may need to be tuned for different acquisitions and species. The log-Jacobian (logJac) images are calculated from the "to-MDT" warps using *CreateJacobianDeterminantImage* with the *UseGeometric* option. All contrasts are smoothed with a 3 voxel sigma Gaussian kernel, using the *SmoothImage* command.

SurfStat, ANTsR, or FSL Randomise are called to provide parametric or non-parametric voxel-based analysis in Stage 8. For parametric statistics two single-tailed t-tests are performed in opposite directions, and statistical maps are generated for the *t*-value, uncorrected *p*-value, and effect size.

5000 permutations are used for nonparametric statistics. Subsequently the multiple-comparison correction is done using False-Discovery Rate (FDR) (Genovese et al., 2002) to produce q-values.

## 2.2. Animals and specimen preparation

Animal procedures were approved by the Duke University Institutional Animal Care and Use Committee. To model epileptogenesis a small-diameter cannula (Plastics One) was stereotactically inserted into the right basolateral amygdala of anesthetized C57BL/6 mice ($n$=10), and KA (0.3 µg in 0.5 µl phosphate-buffered saline [PBS]) was infused at a rate of 0.11 µl/min (Liu et al., 2013). A cohort of 10 control animals was infused similarly with PBS. Twelve weeks following the infusion, the brain specimens were prepared for scanning, as described in (Johnson, 2000; Johnson et al., 2002; Johnson et al., 2007). After being anesthetized to a surgical plane, mice were perfused through the left ventricle with outflow from the right atrium. The blood was flushed out with 0.9% saline at a rate of 8 ml/minute, for 5 minutes. Fixation was done via perfusion with a 10% solution of neutral buffered formalin phosphate containing 10% (50 mM) Gadoteridol (ProHance, Bracco Diagnostics Inc., Cranbury, NJ), at 8 ml/minute for 5 minutes. The heads were removed and soaked in 10% formalin buffer for 24 hours, before being transferred to a 0.01 M PBS solution containing 0.5% (2.5 mM) Gadoteridol at 4 °C for 5–7 days. This reduced the spin lattice relaxation time (T1) of the tissue to $\sim$ 100 ms. Extraneous tissue was removed, and specimens were placed in MRI-compatible tubes, immersed in perfluoro polyether (Galden Pro, Solvay, NJ) for susceptibility matching and to prevent dehydration.

## 2.3. Image acquisition and post-processing

All specimens were scanned on a 7-Tesla small animal imaging system equipped with an Agilent VnmrJ 4 console. A custom silver solenoid coil ($d$ =13 mm) was used for RF transmission and reception. MR-DTI images were acquired using a 3D diffusion-weighted spin-echo sequence with repetition time (TR)=100 ms, echo time (TE)=14 ms, and b-value=1600 s/mm$^2$. The image array size was 400x200x160, over a 20.0x10.0x8.0 mm field of view, producing 50 µm isotropic image resolution. The diffusion sampling protocol included 6 diffusion directions (Jiang and Johnson, 2010) and 1 non-diffusion-weighted (b0) measurement. Total acquisition time was 7 hours. After registering all individual DWI images (each sensitized to a different diffusion direction) to the b0 image with an affine transform we used the Diffusion Toolkit (Wang et al., 2007) to estimate the diffusion tensor and calculate the mean diffusion-weighted image (DWI), axial diffusivity (AD), fractional anisotropy (FA), mean diffusivity (MD), radial diffusivity (RD), and apparent diffusion coefficient (ADC). The DTI parametric images were used as the input for the VBA pipeline.

### 2.4. VBA processing

To examine the effect of template construction we ran the VBA pipeline for two scenarios (controls vs. phantoms; and control vs. KA-injected animals), using 12 different registration parameters sets, and two template generation strategies, for a total of 48 times. The first strategy used only the control animals to construct the MDT (denoted as "C"). In the second strategy all the animals contributed to the MDT ("A"), similarly to (Avants et al., 2010). For a given set of registration parameters, both the phantom and KA runs used the same MDT for "C", while for "A" a new MDT needed to be generated with each pipeline run.

We ran the first three stages of the pipeline only once, because it was not until Stage 4 that any parameters were varied. For these common stages, we used the Waxholm Space mouse brain atlas (Johnson et al., 2010; Calabrese et al., 2015b) to provide the orientation for rigid registration. A native image from the study, padded along y with 12 voxels, defined the reference space. Thus, the final array size was 400x212x160 with 50 µm isotropic resolution. One control subject was arbitrarily selected as the target for affine registration. For both the rigid and affine stages, DWI images were registered with a gradient step of 0.1 voxels, using the Mattes similarity metric (32 bins, 1e-8 convergence threshold, 20-iteration convergence window). Registration was constrained to two down-sampled levels of 6x and 4x, with a maximum of 3000 iterations, with smoothing sigmas of 4 and 2 voxels, respectively. *Histogram matching* and *estimate learning rate once* options were used.

The pipeline runs diverged at Stages 4 and 5, where we varied the three SyN-specific parameters required by *antsRegistration*. The gradient step size, referred to here as the singular "SyN" parameter, took on values of 0.1, 0.25, and 0.5 voxels. The regularization parameter for the velocity ("*update*") field ("RegU") assumed values of 3 and 5. Finally, the regularization parameter for the *total* warp field ("RegT") took on values of 0 and 0.5 voxels. Thus the parameter space of $MDT(SyN, RegU, RegT)$ was 2x3x2x2 = 24 permutations. In the absence of metrics to guide our selection, our "best-guess" was $C(0.25, 3, 0,5)$.

The FA images were used to drive all pair-wise registrations via the cross correlation (CC) similarity metric with a 4-voxel radius and dense sampling. The convergence threshold and window were the same as in previous stages. We used 4 sampling levels, 8x, 4x, 2x, and 1x, with a maximum of 4000 iterations each; with smoothing sigmas of 4, 2, 1, and 0 voxels, respectively. A 3-voxel radius was used for smoothing the images, before voxel-wise statistical analysis.

*2.5. Temporal performance of the pipeline*

To examine computational efficiency we simulated the runtime of the $C(0.25,3,0.5)$ KA analysis when using a high-end workstation and the cluster with 1-6 nodes ($n_{nodes}$). We compared the runtimes for the 24 registration parameters sets when using 4 nodes. In practice, the desired number of multi-threaded registration jobs were assigned to a given node by requesting the appropriate integer fraction amount of memory when calling Slurm, while allowing them to share the cores on that node.

We evaluated: 1) the real ("wall-clock") time for each cluster job; 2) its corresponding total CPU time (processing time of the workload normalized to one processor); 3) a conversion factor relating the two; and 4) the distribution of jobs across $n_{nodes}$ during a given Stage. Slurm's *sacct* command provided the first two quantities via its *CPUTime* and *TotalCPU* fields. From this, we estimated the *CPUTime*/*TotalCPU* conversion factor to be 0.0325 ± 6.6e-04, very close to the theoretical limit of 1/32 (0.0312) for 16 hyper-threaded cores Lastly, given that jobs are to be distributed evenly across nodes, the lists of jobs for each node were easily determined. Each node's workload was calculated by summing the *TotalCPU* of all its jobs, and converted to real time (*CPUTime*). A Stage's runtime was taken to be the longest of these *CPUTimes*. We only considered the jobs from Stages 4 and 5, since these rate-limiting steps serve as an excellent proxy for the temporal performance of the entire pipeline. The combined Stage 4 and Stage 5 runtimes, sorted by constant parameter groups, were log10 transformed to improve normality and to illustrate relative changes in compute efficiency, before performing paired *t*-tests. Resulting effect sizes are thus reported as runtime multipliers.

To compare temporal performance between a workstation and the cluster we calculated a conversion factor based on the average iteration time over the same 3 randomly selected SyN registration jobs. The 3 jobs were run in parallel on a single cluster node, and in serial on our most powerful workstation (12 cores [24 hyper-threaded] x 2 Intel Xenon E5-2650). We chose serial processing on the workstation because the ability to run parallel jobs is RAM limited.

## 2.6. Manual labels and Dice coefficients

An atlas based segmentation using a novel symmetrized atlas featuring 166 regions on each side, and having as initial point the parcellations described in (Calabrese et al., 2015b) provided label sets for 5 KA brains. The automated labels were generated using $C$(0.5,3,1) and provided the starting point for manual corrections. Four regions, left/right hippocampus (Hc) and left/right caudate-putamen/striatum (CPu), were then manually delineated. The same person (RJA) performed all segmentations using Avizo (FEI, Burlington, MA); and multiple contrasts (AD, MD, and RD).

Once each pipeline run completed, the resulting label set was used to calculate Dice coefficients, the "silver standard" for evaluating spatial registration (Avants et al., 2011). These were generated via the ANTs *LabelOverlapMeasures* command. Ipsilateral to the injections, the right Hc Dice values characterize how well atrophy was recovered. The left CPu functioned as a control, as it was minimally impacted by the KA. The left Hc and right CPu were pseudo-controls, featuring structural correlation with the right Hc. An in-depth analysis of the Dice coefficients is in the Supplemental Material. There, the values from the same subject were paired, such that they had 3 of the 4 registration/MDT parameters in common. This kept all variables constant except for one, and its effect could be measured using paired *t*-test across all combinations of constant parameters and subjects. For SyN, three separate *t*-tests were performed, (0.1 > 0.25), (0.1 > 0.5), and (0.25 > 0.5). For each t-test, $n_{pairs}$ = 60 (24 parameter sets*5 specimens/2 groups), except for the SyN comparisons, where $n_{pairs}$ = 40 (24 *5 /3).

## 2.7. The validation framework

We propose a framework for evaluating VBA workflows in the small animal brain (**Figure 2**). This is based on simulated morphological changes, and quantifying their subsequent recovery. There are two primary components: morphological phantom creation, and metric calculation based on VBM.

### 2.7.1. Generation of morphological phantom data

Our primary goal in validating the pipeline was to recover the atrophy or hypertrophy induced in our phantoms. Specifically, we induced hypertrophy in the left CPu and atrophy in the right Hc. An asymmetric approach helps better isolate the opposing morphometric changes; and allows for testing whether any software in the pipeline reverses L-R axis. Creating phantoms with atrophic Hc suited our evaluation purposes since we expected hippocampal atrophy in the KA-injected mice. Simulating atrophy in Hc provided insight into the expected performance for the actual KA group.

We generated phantom images (**Figure 3**), with volumetric changes of ~±14%. We started with a set of control images and their corresponding label sets (A), the latter of which had been automatically produced in Stage 6 with the "best-guess" parameters, $C(0.25,3,0.5)$. From these, we used MATLAB to extract subject-specific binary masks corresponding to the left CPu (B, top) and right Hc (B, bottom). The *imdilate* and *imerode* MATLAB commands were used to alter the regions, until they approached the target volume. The CPu mask was dilated and Hc mask was eroded each by one voxel (C). The two modified masks were then recombined into a single target mask. The original masks were merged and diffeomorphically registered to the target mask with registration parameters $(0.5,3,1)$. The MeanSquares image similarity metric was used with full sampling. To illustrate the voxel-wise volumetric change induced across each structure (D) shows the log-Jacobian for the target-to-original warp. Values less than zero represent atrophy, and greater than zero, hypertrophy. The phantom images for each subject were produced by applying the resulting warp to all contrast images with *antsApplyTransforms* using linear interpolation (E). Creating phantoms for 10 subjects took ~14 minutes for 400x212x160 images, when running in parallel on the cluster. We measured the induced volumetric changes through the mean of the Jacobian across the original masks. The average percent change across all subjects was used as the "target" for evaluating performance. A one-voxel dilation of the CPu corresponded to a 13.6±0.8% change in volume (0.128 in terms of log-Jacobian). The one-voxel erosion produced a −13.6±1.7% change in the Hc (−0.146 log-Jacobian).

**Figure 3** illustrates seen how well VBA recovers the induced morphological changes in phantom images. We present here the results for our best-guess SyN parameters, $C(0.25, 3, 0.5)$. While the effect size (E) exhibits what might be considered noise, the majority of this is eliminated when considering voxels with *p*-values below 0.05 (F). As desired, the clusters of statistical significance are largely confined to the input masks.

*2.7.2. Evaluation metrics for phantom analysis*

We propose four quantitative metrics for evaluating performance of the VBA pipeline: the *distance from target*, the sensitivity index, and the Area Under the Curve (AUC) and True Positive Rate (TPR) at *p=0.05* obtained from Receiver Operating Characteristic (ROC) plots. These quantify the accuracy (in sign and magnitude) of the reported effects, the spatial precision, and the expected tradeoff between true and false positives at various statistical thresholds. Although not scalar metrics, ROC plots are included as well.

The *distance from target* is defined as the absolute distance between the simulated volumetric change and the value implied by the effect size, measured in percentage. The effect size, in the native units of the logJacobian, is averaged across each structure in which we have induced change. The implied volume change in percent is: $\Delta V_{implied} = ((exp(effect_{structure})\text{-}1) \times 100\%$. Consequently, the distance from target is: $\Delta d = |\Delta V_{simulated} \text{-} \Delta V_{implied}|$. Only the absolute value $|\Delta d|$, is considered here, although in some cases it may be of interest if the volumetric change is being overestimated. It is desirable to minimize this total distance as it indicates higher accuracy, and thus we plot it with the *y*-axis inverted.

To quantify the localization of the induced effects we used a sensitivity index, *d'* ("d-prime") (Green and Swets, 1966), and looked for "effect leakage", nearby falsely significant effects primarily arising from bias related to the model priors employed during spatial normalization. We treat the effect size as a signal and the sensitivity index is: $d' = (\mu_S - \mu_N)/\sqrt{(\sigma^2_S + \sigma^2_N)}$. Here, $\mu_S$ and $\sigma_S$ are the mean and standard deviation of the signal, and $\mu_N$ and $\sigma_N$ of the noise, with d-prime indicating how readily a signal can be detected. With perfect registration, the effect size within an altered structure ("signal") should be easily distinguished from effect size immediately outside of the structure ("noise"). We create binary masks for the noise, referred to as the leakage regions, by dilating the masks of the altered structures by two voxels, then removing the original generating structure and any voxels that belong to neighboring structures that have also been altered. Likewise, an inner shell for each structure is created to approximately match the volume of the corresponding adjacent leakage region. The effect size of this region is considered rather than the entire structure. To estimate d-prime we measured the distribution of the effect size within the inner shell and within its leakage region.

Determining whether a voxel is significant is binary classification task based on a threshold, and can be characterized by an ROC. We constructed ROCs based on $p$- and $q$-values. Ideally, all voxels within an altered structure would be significant (True Positives), but none outside (False Positives). For a given threshold, the TPR is the fraction of significant voxels within the structure. The False Positive Rate (FPR) is the amount within the brain, but outside the structure. An ROC is constructed by plotting the FPR along the x-axis and the TPR along the y-axis; and the AUC is calculated by approximating the area with finite trapezoids.

Each metric was calculated for the 24 parameter sets for the right Hc (atrophy) and the left CPu (hypertrophy), and then group-/pair-wise sorted in the same fashion as the Dice coefficients by varying 1 of the 4 parameters at a time. Similarly, paired $t$-tests were performed across the constant parameter groupings for these 4 phantom metrics and the average Dice coefficients, and the $p$-values and median effect sizes were recorded. Based on these pairwise comparisons, we selected two scenarios for side-by-side comparison of the KA VBA results, in which the impact of varying one parameter at a time—RegT and SyN in this case—was visually apparent. More scenarios, including the variation of RegU and MDT group, are included in the Supplemental Material; together with the results for the SyN(0.1 > 0.5) and SyN(0.25 > 0.5) $t$-tests, and scatter plots examining the correlation between the Dice coefficients and the phantom metrics.

To produce an average phantom ranking the 24 parameter sets were ranked according to each metric, and ranks were averaged. The average phantom and Dice ranking, as well as the runtimes were computed - to integrate all metrics available. The KA VBA results corresponding to the extremes and the median of the ranked results were compared, to illustrate the variation in unguided VBA.

## 3. Results

To address preclinical imaging needs, we have developed a cluster-based VBA pipeline for small animal multivariate brain analysis, SAMBA. We propose a VBA validation framework consisting of morphological phantoms and VBA-specific metrics. We have used SAMBA in a thorough evaluation of time efficiency gained from parallel processing, and applied it to a model of epilepsy - illustrating the wide effects of parameter choices on VBA, and how phantoms can inform parameter's selection.

### 3.1. Temporal performance

A major advantage of HPC is the increased throughput. **Figure 4A** shows the runtimes for Stages 4 and 5 using $C(0.25,3,0.5)$ for a single workstation, as well as 1-6 cluster nodes. Compared to serial job scheduling on a workstation with a similar processor (first bar), we noted a speedup of ~2.11 by moving to the cluster. The ability to run parallel jobs with high memory requirements is a clear advantage of HPC, even if only using one node. Each additional node increases this factor by ~0.8. Adding nodes decreased the total runtime, approaching the lower limit of $1/n$. Using 6 nodes, the VBA time decreased from ~1.5 weeks to ~1.5 days, an 86% increase in efficiency.

The VBA pipeline runtimes ranged from 7.3 hours for $C(0.5,5,0)$ to 187 hours (7.8 days) for $A(0.1,3,0.5)$. The largest effect of changing any one parameter was attributed to SyN (B). SyN(0.1) runs typically took 4x longer than SyN(0.25). Using RegU(5) instead of RegU(3) resulted in a ~25% reduction in runtime (C). Similarly, a modest difference (~30%) came from choosing RegT(0.5) (D).

The ~45% increase in runtime when creating an MDT from all the subjects was consistent across comparisons (E). In conclusion, the time penalty was high when SyN was small, or even when SyN was modest and RegU was small. This suggests that better performing parameter groups require longer runtimes, and a runtime penalty must be taken into account when selecting parameters.

### 3.2. Evaluation of processing parameters via phantom VBA

Kainic acid Dice values in the Hc (both left and right) ranged from 86.66% to 95.49%, with a mean of 92.58±1.35%. Across the CPu, the Dice ranged from 88.89% to 95.20%, with a mean of 93.46±1.26%. Subject-wise paired *t*-tests are tabulated and visualized in **Supplemental Table S1** and **Figure S1**, respectively. **Table 1** illustrates the impact of each parameter choice.

| Parameter Comparison | Dice | | \|Δd\| | | d' | | AUCx100 | | TPR@p=0.05 | |
|---|---|---|---|---|---|---|---|---|---|---|
| | Hc | CPu | Hc | CPu | Hc | CPu | Hc | CPu | Hc | CPu |
| **SyN: 0.1>0.25** | | | | | | | | | | |
| effect size: | 0.42% | 0.27% | 0.18% | 0.17% | **0.143 | **0.168 | **0.09 | **0.35 | **9.5% | **24.5% |
| *p*-value: | 0.069 | 0.251 | 0.028 | 0.071 | 0.002 | 0.002 | 2.3e-04 | 1.4e-04 | 1.3e-4 | 7.3e-04 |
| **SyN: 0.1>0.5** | | | | | | | | | | |
| effect size: | 0.52% | 0.37% | 0.30% | 0.05% | **0.148 | **0.180 | **0.13 | **0.46 | **11.3% | **31.5% |
| *p*-value: | 0.049 | 0.064 | 0.008 | 0.366 | 0.003 | 0.002 | 2.1e-05 | 8.8e-05 | 8.2e-05 | 7.6e-04 |
| **SyN: 0.25>0.5** | | | | | | | | | | |
| effect size: | 0.16% | **0.16% | 0.10% | -0.16% | 0.016 | 0.031 | **0.03 | 0.08 | **2.3% | 4.1% |
| *p*-value: | 0.110 | 6.3e-04 | 0.212 | 0.102 | 0.189 | 0.016 | 0.0025 | 0.048 | 0.002 | 0.015 |
| **RegU: 5>3** | | | | | | | | | | |
| effect size: | −0.01% | −0.08% | **−0.14% | −0.21% | **-0.085 | **−0.070 | **−0.07 | **−0.19 | **−4.8% | **−0.2% |
| *p*-value: | 0.418 | 0.688 | 4.4e-04 | 0.019 | 1.3e-05 | 8.2e-05 | 1.5e-04 | 2.3e-04 | 1.9e-04 | 3.6e-04 |
| **RegT: 0.5>0** | | | | | | | | | | |
| effect size: | **0.93% | **0.82% | 0.1% | **0.31% | **0.159 | **0.142 | **0.22 | **0.85 | **24.0% | **39.8% |
| *p*-value: | 8.3e-04 | 3.1e-04 | 0.158 | 5.8e-04 | 9.1e-07 | 1.9e-05 | 5.3e-08 | 1.3e-09 | 3.1e-09 | 1.3e-09 |
| **MDT: All>Ctrl** | | | | | | | | | | |
| effect size: | **0.83% | 0.03% | **0.14% | −0.05% | 0.013 | **−0.035 | **0.02 | −0.07 | **−1.4% | −2.0% |
| *p*-value: | 8.5e-06 | 0.227 | 4.2e-04 | 0.061 | 0.020 | 3.3e-05 | 0.002 | 0.011 | 0.003 | 0.005 |

**Table 1. The *p*-values and effect sizes of the pairwise *t*-tests of SyN, RegU, RegT, and MDT, for Dice coefficients and phantom metrics.** **indicates effect sizes with *p*-value < 0.005

Overall, the total (deformation) regularization (RegT) and gradient step (SyN) produced the largest effects in both Dice and phantom metrics. RegT paired *t*-tests featured the smallest *p*-values and, in AUC and TPR, effect sizes ~2x greater than the closest values produced by SyN. The update (velocity) field regularization (RegU) had a smaller but significant effect per the phantom metrics, which was not captured by the Dice coefficients. The choice of MDT group had a sizable effect per the Dice coefficients, particularly in the atrophied right KA hippocampi, while the phantom metrics incorrectly did not capture this effect. **Table 1** shows which phantom metric best correlated to the Dice (see also Supplemental **Figure S7**). AUC was most tightly correlated to the Dice with R=0.708, *p* = 1.076e-4 (atrophy) and R=0.836, *p*=3.75e-7 (hypertrophy). By isolating the role of each parameter, the effect size and *p*-value (insets) from a paired *t*-test (**Table 1**) can help identify optimal performers.

      **Figure 5** illustrates the impact of varying RegT on performance metrics (see also the Supplemental **Figures S2**, **S3**, and **S5** for the effects of varying SyN, RegU, and the MDT group). The automated labels of the 24 KA VBA runs are used for calculating average Dice coefficients (A). The metrics based on the 24 phantom VBA runs include: absolute distance from target (B), sensitivity index (C), the Receiver Operating Characteristic (ROC) plot (D), ROC Area Under Curve (E), and ROC TPR at *p* = 0.05 (F). Compared to other parameters RegT(0.5) produced the largest significant effects on Dice coefficients (**Figure 5A**). A trend emerged in |Δ*d*| (B), in which the effects due to changing a given parameter were either highly variable or in favor of values that perform more poorly per other phantom metrics. RegT induced significant variation in all phantom metrics (apart from |Δ*d*|). This is evident in *d'* (C), AUC (E), and TPR (F). We chose Parameter Group 10, (*A*(0.1,5,xx, indicated by the arrows in **Figure 5**) because of its large impact on most metrics, and assessed the corresponding KA results (**Figure 6**). Both atrophy and hypertrophy were more expansive if using RegT(0.5) relative to RegT(0), notably in the contralateral cortex, caudate putamen, hippocampus, and amygdala. Differences were large enough that varying RegT could lead to divergent conclusions.

      Because large effects were detected due to using SyN(0.1) over SyN(0.25), in **Table 1** we examined the SyN effect for KA VBA (**Figure 7**). We chose parameter group *A*(xx,3,0.5) since it featured large differences between the three SyN values across all the phantom metrics (see arrows in Supplemental **Figure S2**). There was little difference in the atrophy detected near the caudate

putamen (yellow slice). However, the extent and localization of the atrophy in the other slices varied with SyN. SyN(0.1) identified larger clusters in the cortex/corpus callosum and periventricular regions.

Volume increases (hypertrophy) in KA treated animals were apparent in the ipsilateral hemisphere, as well as in the contralateral amygdala and the adjacent hippocampus. Moreover, the contralateral caudate putamen showed hypertrophy. In general, the clusters extent increased as SyN decreased, which is consistent with parameters recommended for *antsRegistration* in human data (Tustison, 2013). Similar comparisons of the VBA effects of RegU and MDT group can be found in **Supplemental Figures S4** and **S6**, respectively.

*3.5. Control vs. kainic acid VBA*

**Table 2** ranks all parameter groups according to the 4 phantom metrics. In general, the smaller SyNs, and more total regularization performed better, with $A(0.1,3,0.5)$ as the top performer. For a comparison of the extreme and average cases according to the phantom rankings, the KA VBA results for this parameter set were plotted in **Figure 8** alongside the median case of $A(0.25,5,0.5)$, and the worst performer $A(0.5,5,0)$.

| Parameter Group | \|Δd\| (%) | | d' | | (AUC−0.98)x100 | | TPR@p=0.05(%) | | Phantom Average |
|---|---|---|---|---|---|---|---|---|---|
| | *Hc* | *CPu* | *Hc* | *CPu* | *Hc* | *CPu* | *Hc* | *CPu* | *Average* |
| C(0.1,3,0) | 1.89 *11* | 2.58 *19* | 1.34 *10* | 1.06 *6* | 1.542 *12* | 1.035 *12* | 80.9 *10* | 55.5 *11* | 11.4 *12* |
| C(0.1,3,0.5) | 2.41 *23* | 2.89 *24* | 1.64 *2* | 1.38 *1* | 1.645 *3* | 1.572 *2* | 92.6 *2* | 97.7 *1* | 7.3 *3* |
| C(0.1,5,0) | 1.74 *4* | 2.30 *6* | 1.21 *19* | 0.90 *17* | 1.404 *18* | 0.495 *17* | 65.9 *15* | 30.5 *16* | 14.0 *15* |
| C(0.1,5,0.5) | 2.08 *19* | 2.54 *18* | 1.47 *4* | 1.20 *3* | 1.633 *4* | 1.395 *4* | 90.0 *3* | 92.9 *3* | 7.3 *4* |
| C(0.25,3,0) | 1.79 *7* | 2.41 *9* | 1.23 *16* | 0.91 *16* | 1.377 *19* | 0.516 *16* | 64.0 *17* | 30.6 *15* | 14.4 *16* |
| C(0.25,3,0.5) | 1.90 *13* | 2.49 *16* | 1.38 *7* | 1.08 *5* | 1.593 *6* | 1.248 *5* | 85.8 *5* | 76.4 *5* | 7.8 *5* |
| C(0.25,5,0) | 1.79 *9* | 2.13 *3* | 1.16 *24* | 0.88 *21* | 1.264 *23* | 0.185 *22* | 55.5 *22* | 24.6 *21* | 18.1 *21* |
| C(0.25,5,0.5) | 1.76 *5* | 2.43 *11* | 1.31 *12* | 1.01 *10* | 1.544 *11* | 1.097 *8* | 81.2 *9* | 60.0 *9* | 9.4 *8* |
| C(0.5,3,0) | 1.62 *2* | 2.65 *21* | 1.22 *17* | 0.92 *15* | 1.347 *20* | 0.528 *15* | 61.9 *18* | 26.5 *19* | 15.9 *18* |
| C(0.5,3,0.5) | 1.77 *6* | 2.66 *22* | 1.37 *8* | 1.03 *9* | 1.562 *8* | 1.133 *7* | 83.5 *7* | 63.8 *7* | 9.3 *7* |
| C(0.5,5,0) | 1.59 *1* | 2.48 *13* | 1.17 *23* | 0.84 *22* | 1.255 *24* | 0.258 *20* | 56.0 *21* | 22.3 *22* | 18.3 *22* |
| C(0.5,5,0.5) | 1.68 *3* | 2.42 *10* | 1.30 *14* | 0.99 *12* | 1.528 *14* | 1.054 *9* | 79.5 *12* | 55.9 *10* | 10.5 *9* |
| A(0.1,3,0) | 2.11 *20* | 2.34 *7* | 1.35 *9* | 1.04 *8* | 1.544 *10* | 1.037 *11* | 78.7 *13* | 54.9 *12* | 11.3 *11* |
| A(0.1,3,0.5) | 2.45 *24* | 2.86 *23* | 1.66 *1* | 1.35 *2* | 1.662 *1* | 1.617 *1* | 92.8 *1* | 97.6 *2* | 6.9 *1* |
| A(0.1,5,0) | 1.99 *15* | 2.48 *12* | 1.20 *20* | 0.89 *20* | 1.428 *15* | 0.424 *18* | 64.6 *16* | 27.1 *18* | 15.5 *17* |
| A(0.1,5,0.5) | 2.13 *21* | 2.48 *15* | 1.49 *3* | 1.16 *4* | 1.652 *2* | 1.409 *3* | 89.9 *4* | 92.1 *4* | 7.0 *2* |
| A(0.25,3,0) | 2.15 *22* | 2.00 *1* | 1.21 *18* | 0.90 *18* | 1.418 *16* | 0.299 *19* | 61.1 *19* | 29.5 *17* | 16.3 *19* |
| A(0.25,3,0.5) | 2.03 *17* | 2.48 *14* | 1.44 *5* | 1.04 *7* | 1.615 *5* | 1.202 *6* | 85.0 *6* | 73.4 *6* | 8.3 *6* |
| A(0.25,5,0) | 1.90 *12* | 2.35 *8* | 1.17 *21* | 0.83 *24* | 1.338 *21* | 0.143 *23* | 55.2 *23* | 20.0 *24* | 19.5 *23* |
| A(0.25,5,0.5) | 1.87 *10* | 2.44 *12* | 1.34 *11* | 0.97 *13* | 1.557 *9* | 1.029 *13* | 80.3 *11* | 54.4 *13* | 11.5 *13* |
| A(0.5,3,0) | 2.06 *18* | 2.18 *5* | 1.25 *15* | 0.90 *19* | 1.417 *17* | 0.238 *21* | 57.1 *20* | 26.3 *20* | 16.9 *20* |
| A(0.5,3,0.5) | 1.92 *14* | 2.62 *20* | 1.39 *6* | 0.99 *11* | 1.570 *7* | 1.052 *10* | 82.1 *8* | 60.0 *8* | 10.5 *10* |
| A(0.5,5,0) | 2.00 *16* | 2.13 *4* | 1.17 *22* | 0.84 *23* | 1.295 *22* | 0.008 *24* | 52.9 *24* | 21.2 *23* | 19.8 *24* |
| A(0.5,5,0.5) | 1.79 *8* | 2.51 *17* | 1.30 *13* | 0.93 *14* | 1.529 *13* | 0.926 *14* | 77.2 *14* | 46.8 *14* | 13.4 *14* |

**Table 2. The 4 phantom metrics of VBA pipeline evaluation, relative rank (italics), and average phantom rank (bold).** *C*=only control group used for MDT, *A*=all individuals used for MDT, with (SyN, RegU, RegT).

**Table 3** contains the results of the phantom rankings with the Dice coefficients, pipeline runtimes, the average phantom ranking from **Table 2**, and a ranked average of the three. This combined ranking favored larger steps sizes and smaller MDT groups due to faster convergence, and ranked *C*(0.25,5,0.5) as the top performer. Among the MDT(All) groups, *A*(0.1,5,0.5) was the top, and *A*(0.1,3,0) was the median performer.

| Parameter Group | Dice % - 90% | | Average Dice | | Runtime (Hours) | | Phantom Ranking | Dice + Runtime + Phantom Average | |
|---|---|---|---|---|---|---|---|---|---|
| | Hc | CPu | | | | | | | |
| C(0.1,3,0) | 2.66 _12_ | 3.23 _10_ | 11 | _11_ | 85.0 | _19_ | _12_ | 14.0 | _13 (tie)_ |
| C(0.1,3,0.5) | 2.51 _17_ | 2.99 _14_ | 15.5 | _16_ | 134.0 | _23_ | _3_ | 14.0 | _13 (tie)_ |
| C(0.1,5,0) | 2.37 _20_ | 2.95 _16_ | 18 | _17_ | 33.2 | _15_ | _15_ | 15.7 | _19_ |
| C(0.1,5,0.5) | 2.75 _10_ | 3.37 _5_ | 7.5 | _7_ | 128.5 | _22_ | _4_ | 11.0 | _10_ |
| C(0.25,3,0) | 1.26 _22_ | 2.48 _20_ | 21 | _22_ | 11.2 | _6_ | _16_ | 14.7 | _16_ |
| C(0.25,3,0.5) | 2.71 _11_ | 3.44 _2_ | 6.5 | _5_ | 54.3 | _17_ | _5_ | 9.0 | _7 (tie)_ |
| C(0.25,5,0) | 1.24 _23_ | 2.49 _18_ | 20.5 | _21_ | 12.4 | _9_ | _21_ | 17.0 | _22_ |
| C(0.25,5,0.5) | 2.76 _9_ | 3.41 _4_ | 6.5 | _6_ | 8.6 | _3_ | _8_ | 5.7 | _1_ |
| C(0.5,3,0) | 0.20 _24_ | 2.29 _24_ | 24 | _24_ | 9.9 | _4_ | _18_ | 15.3 | _17 (tie)_ |
| C(0.5,3,0.5) | 2.56 _14_ | 3.27 _8_ | 11 | _12_ | 10.0 | _5_ | _7_ | 8.0 | _2 (tie)_ |
| C(0.5,5,0) | 1.41 _21_ | 2.48 _19_ | 20 | _20_ | 7.3 | _1_ | _22_ | 14.3 | _15_ |
| C(0.5,5,0.5) | 2.51 _16_ | 3.20 _11_ | 13.5 | _14_ | 7.5 | _2_ | _9_ | 8.3 | _4 (tie)_ |
| A(0.1,3,0) | 3.26 _7_ | 3.25 _9_ | 8 | _8_ | 111.5 | _21_ | _11_ | 13.3 | _12_ |
| A(0.1,3,0.5) | 3.34 _5_ | 3.13 _13_ | 9 | _9_ | 187.2 | _24_ | _1_ | 11.3 | _11_ |
| A(0.1,5,0) | 3.18 _8_ | 2.98 _15_ | 11.5 | _13_ | 48.5 | _16_ | _17_ | 15.3 | _17 (tie)_ |
| A(0.1,5,0.5) | 3.58 _1_ | 3.43 _3_ | 2 | _2_ | 90.5 | _20_ | _2_ | 8.0 | _2 (tie)_ |
| A(0.25,3,0) | 2.59 _13_ | 2.59 _17_ | 15 | _15_ | 19.1 | _14_ | _19_ | 16.0 | _20_ |
| A(0.25,3,0.5) | 3.51 _2_ | 3.48 _1_ | 1.5 | _1_ | 69.6 | _18_ | _6_ | 8.3 | _4 (tie)_ |
| A(0.25,5,0) | 2.53 _15_ | 2.46 _22_ | 18.5 | _18_ | 18.0 | _13_ | _23_ | 18.0 | _23_ |
| A(0.25,5,0.5) | 3.40 _3_ | 3.34 _6_ | 4.5 | _3_ | 13.7 | _11_ | _13_ | 9.0 | _7 (tie)_ |
| A(0.5,3,0) | 2.40 _18_ | 2.47 _21_ | 19.5 | _19_ | 13.5 | _10_ | _20_ | 16.3 | _21_ |
| A(0.5,3,0.5) | 3.35 _4_ | 3.32 _7_ | 5.5 | _4_ | 14.8 | _12_ | _10_ | 8.7 | _6_ |
| A(0.5,5,0) | 2.39 _19_ | 2.37 _23_ | 21 | _23_ | 12.0 | _8_ | _24_ | 18.3 | _24_ |
| A(0.5,5,0.5) | 3.33 _6_ | 3.16 _12_ | 9 | _10_ | 11.5 | _7_ | _14_ | 10.3 | _9_ |

**Table 3. Dice, runtimes, average phantom rankings, and their combined average rankings, for the 24 parameter sets.**

Informed by phantom studies, we examined the variations in VBA results for the KA-injected mice (**Figure 8**). Results have been ordered from left to right according to their phantom rankings, the best on the left. It is apparent that VBA is highly sensitive to non-linear registration parameters. The significant contralateral regions of hypertrophy covered fewer and smaller areas with poorer registration performance, and there were hardly any significant voxels in the hippocampus and amygdala for $A$(0.5,5,0). The median parameter set detected much of the hypertrophy, but not to the extent of that of $A$(0.1,3,0.5), and largely missed the cluster in the caudate putamen. This can be summarized as a consistent increase of the number of false negatives and underreporting of treatment effects.

The atrophy observed for the best performer was mostly ipsilateral to the injection site, localized to the hippocampus, hypothalamus, cingulate cortex, primary somatosensory and temporal association cortex, as well as the caudate putamen, globus pallidus and thalamus (anterodorsal, ventral, reticular nuclei). Contralateral atrophy was also noted, to a smaller extent, in the medial geniculate, hypothalamus, temporal association cortex, and the periventricular hippocampus. Approximately a third of these regions would not be reported as significant based on the median parameter set, and another third would be overlooked by the poorest performer. In both scenarios, the choice of SyN parameters had a considerable impact on the VBA conclusions.

## 4. Discussion and Conclusions

Phenotyping rodent models of neurological and psychiatric conditions poses substantial challenges because number and size of the images to be analyzed. Image analysis pipelines aim to provide quantitative image-based biomarkers, in a reproducible and automated manner, while meeting the needs for accuracy and efficiency. Previous efforts have been largely dedicated to automating pipelines for human brain images, and several efforts have been made for rodent brain images (Ad-Dab'bagh et al., 2006; Sawiak et al., 2009a). Existing methods for evaluating VBA only capture aspects of the processing chain. Here, we present SAMBA, a VBA pipeline for the rodent brain with an HPC implementation, and an unprecedented extensive validation effort. HPC resources were used to produce 24 variations of VBA in a mouse model of epilepsy, to identify the most reliable results. Our validation framework is based on simulated atrophy/hypertrophy phantoms. Combined, our work enables timely preclinical VBA with increased confidence.

*4.1. Comparison to previous work*

Pagani et al. (2016) described a VBA pipeline for rodent brain MRI, which relies on ANTs, and features segmentation, label-based analysis, cortical thickness, and VBA. Our approach handles larger and multivariate image sets, using 7 derived DTI contrasts. Because our images have almost 2 times higher resolution, our arrays of 512x256x256 voxels are 6 times larger. To meet the computational demands of high-resolution image analysis we designed our pipeline for an HPC environment A defining feature of our pipeline is that it provides the needed code infrastructure to run in an HPC environment.

A second distinctive feature is the proposed validation framework. To the best of our knowledge, a complete validation framework does not exist for preclinical VBA. However, several aspects of the current framework have historical foundations. Freeborough and Fox (1997) used the Boundary Shift Integral to simulate volumetric change. To achieve more anatomically realistic morphological changes Camara et al. (2006) used physical tissue models in conjunction with finite-element analysis; while Karaçali and Davatzikos (2006) preserved topology by constraining the Jacobians of the deformation fields. These methods are best suited for higher-level mammalian brains that feature extensive folding, which interfaces directly with cerebral-spinal fluid (CSF). For the mouse, VanEede et al. (2013) used Jacobian regularization to simulate both atrophy and hypertrophy. While this resulted in Jacobians that were more uniform and spatially constrained, ours has the advantage of requiring substantially less computation time.

A potential limitation of our work is that the volume changes induced in our phantoms (~14%) may not have been enough to emulate the large deformations in the KA study. On the other hand, it may be more difficult to recover changes of ~10% or less. A future task is to establish a method to quickly produce custom volume changes.

A key component of the validation framework comes from the evaluation metrics. Most often the accuracy for spatial normalization is quantified by label overlap metrics such as Dice or Jaccard coefficients (Avants et al., 2011), or label "entropy" based on lower order tissue segmentations (Robbins et al., 2004). However, these do not fully capture the entire VBA process. More appropriate for VBA, Shen et al. (2007) looked at the number of voxels in which they had induced atrophy, and measured the difference between this target and the number of significant atrophy voxels recovered. Similarly VanEede et al. (2013) compared simulated and recovered atrophy/hypertrophy via Deformation Based Analysis (DBA), to measure the number of true and false positives. While the latter two served as inspiration for some of the metrics we employed, we provide multiple quantitative metrics, and note that our absolute distance to target is based on the effect size, as inherited from SurfStat VBA (Worsley et al., 2009), which is not normalized by the standard deviation.

The sensitivity index has similarities to past work, e.g. VanEede et al. (2013) showed that excluding the significant voxels in an $r=3$ voxel shell surrounding a regions of interest, could eliminate most false positives. This shell is in principle equivalent to our leakage region. Rather than omitting these "almost-correct" voxels, we have used the effect sizes to quantify the precision of spatial normalization and the effect of the smoothing.

*4.2. Temporal performance*

While running parallel jobs across multiple nodes reduces computation time, the need for wise resource management remains. In the simplest case, each job is distributed to one node, which is not efficient (Figure 4). In our preclinical studies requiring 10–25 concurrent jobs, 3 nodes provided a good balance between runtime and efficient resource management. Surprisingly, using 4 nodes can take *longer* than using 3 (**Figure 4**A) if two or more particularly demanding jobs were assigned to the same node. Distributing such jobs to even out resource demands can significantly improve efficiency. Figures 4C and D show large discrepancies in runtimes for smaller SyN values, indicating that some registration jobs converged slowly, or suffered from oscillations with little improvement in quality. To circumvent this, one can limit the iterations at the fully sampled level to ~60 or less. Such strategies can reduce VBA runtime significantly—and are the subject for future work.

While in principle, computational expenses are relatively cheap compared to the cost for producing animal models and the imaging equipment acquisition, maintenance and operation. In practice, analysis takes oftentimes the longest amount of time in an experiment. With our efforts, we try to balance the computational time required to reach a conclusion at the end of an experiment.

### 4.3. Dice coefficients and phantom VBA metrics

Due to the computational efficiency of the HPC implementation, we were able to explore a wide parameter space. This provided insight into the value of phantom metrics, and the importance of registration/MDT choices. Dice coefficients captured the benefit of using the All MDT groups, which the phantom metrics largely failed to do. However, apart from detecting the impact of using RegT(0.5), the Dice did not provide much direction on which registration parameters to use.

$|\Delta d|$ had a non-linear relationship with the choice of parameters, and did not detect significant differences in both the atrophic and hypertrophic cases. In general, the best aggregate performing parameters saw poor performance according to this metric. This is due in part to the demand for capturing a spectrum of diffeomorphic changes with a single set of registration parameters, and the various forms of smoothing occurring throughout the pipeline. A $|\Delta d|$ penalty in accuracy was incurred for improved spatial and statistical sensitivity (*d'* and ROC metrics, respectively). While it is not recommended to use $|\Delta d|$ for optimization, it remains a vital piece of information to be shared alongside VBA results, as it estimates the error in effect sizes.

In contrast to $|\Delta d|$, *d'* was sensitive to virtually any change in the parameters. Assuming that spatial localization is given priority over effect size, it is appropriate to include *d'* in any VBA tuning/optimization. This provides an idea of the uncertainty associated with the spatial extent of observed effects. Further, it can characterize the "VBA SNR" of the system. Note that $d'$ depends of the spatial smoothing kernel, an aspect which we did not explore.

The ROC metrics were sensitive to parameter changes, even more than $d'$, and typically had the smallest $p$-values. They were more sensitive to hypertrophy, as their variance was considerably wider here than in the atrophic scenarios. Like $d'$, these metrics are appropriate candidates for VBA tuning. Beyond adjusting the processing parameters, the ROC provides motivation for a particular $p$- or $q$-threshold. Using this to estimate the TPR and FPR in the real data, one can choose where on the curve to report results, depending on which type of Error (Type 1 vs. Type 2) is more tolerable. Including phantom ROCs should increase the level of transparency and confidence in preclinical VBA results, and will hopefully contribute to wider-spread adoption.

While Dice coefficients are an established standard, they are obtained through labor-intensive manual editing, susceptible to bias, and impractical for "everyday use". Of the phantom metrics, the best candidate for a Dice substitute was the AUC, the two having a high correlation. However, comparing MDT(Controls) and MDT(All) revealed an important shortcoming of using phantoms, i.e. the dependence on the induced volume change. We note that the ~14% volume change was substantially less than what we encountered in the KA data, and it is critical to include this when reporting phantom metrics. It is possible that other metrics would correlate strongly with Dice, had larger volumetric changes been induced in the phantoms. Future phantoms can be tuned to better simulate the data in question, and would require a more sophisticated model beyond the linear expansion/contraction method used here.

*4.3. Selecting registration and MDT parameters*

Apart from which MDT group to use, the phantom metrics gave clear insight into which registration parameter values are more likely to give high quality results: SyN(0.1), RegU(3), and RegT(0.5). As a general application, the phantom metrics could aid in selecting between a limited number of parameters. For example, one may already be confident that SyN(0.3) balances quality of results and runtime, but may want to tune RegT to find the value predicted by ROC metrics to provide the highest TPR/FPR ratio. Sharing such tuning procedures and the relevant phantom metrics will help build the experience of the community. Currently, it seems difficult to find detailed registration parameters reported, much less a justification for their choice and the implications for interpretation of VBA results.

The effect sizes of each parameter on the performance metrics and runtimes can inform the decision on how to get reliable results in a reasonable timeframe. The rankings of **Table 2** are a step towards incorporating such results into a cost-benefit analysis. Instead of simply choosing the highest ranked parameter group according to the phantom metrics (and Dice, if available), it may be wise to take into account that even though $A$(0.1,3,0.5) promises the results with the highest fidelity, it also required the longest time (~1 week). By weighting the Dice and phantom metrics against the runtime, one can get a more balanced sense of "value", particularly if access to high-powered computing resources is limited. Such a weighted ranking is included in **Table 3**, and indicates that $C$(0.25,5,0.5) can deliver results in the upper third of quality, in under 9 hours. Another strategy might be to recognize a mismatch between the expected effect size and runtimes. An obvious example here is RegT(3), which provides benefits in quality that make it worth the ~1/3 increase in time. Although the phantom metrics did not elucidate the benefit of using all subjects to construct the MDT observed in the KA VBA results, the pairwise temporal analysis revealed that one can expect it to take 50% longer—and that such a sacrifice is a low (and predictable) price to pay for the benefit of a minimally-biased template.

## 4.4. Kainic acid VBA

We detected atrophy in the hippocampus, and amygdala near the injection site, and also in the striatum, thalamic nuclei (e.g. the geniculate bodies, zone incerta, and laterodorsal nucleus).

Changes in the ipsilateral hippocampus, striatum, pallidum and thalamus have been well documented in patients with temporal lobe epilepsy (Dreifuss et al., 2001). This study also reported contralateral atrophy in these structures. Of these we detected atrophy in the contralateral thalamus and periventricular hippocampus. We also detected widespread contralateral hypertrophy. There is evidence of contralateral hypertrophy in rodent brains under similar circumstances (Pearson et al., 1986; Dedeurwaerdere et al., 2012). These can be explained by hippocampal neurogenesis (Parent et al., 1997), mossy fiber sprouting (Wuarin and Dudek, 1996), astrogliosis (Li et al., 2008)——and could obscure the VBA detection of atrophy due to neuronal cell death (Altar and Baudry, 1990).

To validate VBA results with histology (**Figure 9**) we examined the hippocampus of a KA injected mouse a PBS injected control. Neurons and astrocytes were visualized using a Leica TCS SL confocal microscope, after staining with antibodies against neuronal nuclei (NeuN, Millipore) and glial fibrillary acidic protein (GFAP, Sigma). The histology revealed neurodegeneration and astrogliosis in KA injected animals.

The choice of registration parameters impacted the detection of brain phenotypes, highlighting the need for mindful VBA. This was evident when varying RegT, where we noted the potential for divergent interpretations. Similar variations in sensitivity may arise if working at a fixed statistical threshold, and if the dataset under consideration has similar variability to ours. One might select registration parameters based on intuition, and our "best-guess" of $C(0.25,3,0.5)$ was not far from the median performer $A(0.25,5, 0.5)$. **Figure 8** shows modest variations between $A(0.25,5,0.5)$ and $A(0.1,3,0.5)$ in the atrophic ipsilateral regions, particularly the hippocampus, and also the striatum, pallidum, cingulate cortex, thalamus and hypothalamus. There was however more variation in the contralateral hypertrophy in the amygdala, cortex, striatum, and hippocampus. Hypertrophy largely vanished when using poorer performing parameter sets e.g. $A(0.5,5,0)$. This may be a compensatory mechanism for severe atrophy in one hemisphere, under the constraint that volume needs to be preserved when all brains are mapped into the same template space.

The overall variability between registration parameter sets underscores the importance of a method for validating VBA, to protect against conforming the results or their interpretation to a preconceived bias. This translates into a need to develop quantitative tools for informing VBA, not only on registration parameters, but also on statistical thresholds and smoothing kernels (Jones et al., 2005). Such tools should allow decisions to be made using a consistent framework, imbuing confidence to researchers, and their audience. The phantoms and the evaluation methods we proposed are starting points for such a toolbox/framework.

## 4.5. Future work

We have applied the pipeline in its entirety, or as independent modules to phenotyping live (Badea et al., 2016), or fixed mouse (Badea et al., 2007; Badea et al., 2012), rat (Calabrese et al., 2013), and primate brain images (Calabrese et al., 2015a). Our future efforts are motivated by the desire for efficient and reliable voxel-based analysis, which addresses an unmet need for validating and sharing VBA results in preclinical MRI. To realize this vision, we need to identify the minimum quantitative validation requirements to become standard in future VBA studies. It would be beneficial to standardize workflows for generating data sets with a range of simulated atrophy and hypertrophy. There is a need for comprehensive, well-characterized evaluation metrics. Phantoms can guide the VBA processing and interpretation of real data. We should next extend the phantom concept beyond VBM, to other contrasts. We have optimized critical parameters and note that there is potential for more efficient algorithms. The effects of other parameters, e.g. the size of the smoothing kernel, need to be more thoroughly investigated. Future work might also consider validation models that employ biologically relevant deformations, a greater range of scales, landmark distances, or region-wise overlaps, as in (Tustison, 2013). While we have incorporated options for both parametric and nonparametric statistics, we have only explored the first case here, and more can be done varying the options for statistical analyses. A deeper consideration of preclinical study design—from data collection to analysis strategy, and statistical modeling—is warranted due to the potential to improve inference from preclinical to clinical studies. Also, a fully determined BIDS standard for small animal imaging and derived data is still work in progress.

In conclusion, it is clear that parallelizing tasks such as image registration and statistical analysis (in particular permutation based nonparametric tests) are worth the effort. Yet, this is not yet widely-adopted, in part because of the upfront effort required for such implementations. We shared our experience in the context of small animal brain image analysis, using a local cluster. Further developments should address cloud portability. While we focused on the brain, such efforts are translatable to other organs (such as heart and lungs), and other species (rats, non-human primates). Lastly, we argue that validation efforts have not received sufficient attention in preclinical VBA, and we propose an evaluation framework, also easily adaptable to other organs, and species.

*4.6. Conclusions*

We addressed the demands of preclinical VBA with an automated pipeline in a local HPC cluster. We identified a need for optimization and validation tools. To address this, we proposed several evaluation metrics to be used in conjunction with phantoms featuring simulated atrophy and hypertrophy. We applied these tools to illustrate how widely VBA results can vary with different registration parameters, using as an example a mouse model of epilepsy. Using such tools, we are able to increase the confidence in VBA results, and quantitatively communicate this confidence. The community shall benefit from further development of a robust evaluation framework for preclinical VBA studies, whether these are performed in local computing environments, university/company HPC resources, or in the cloud. Code repositories are freely available at https://github.com/andersonion/VBA_validation_framework (phantom generation, evaluation metrics); and https://github.com/andersonion/SAMBA (for the VBA pipeline).

**Acknowledgments**

The Duke University Center for In Vivo Microscopy is an NIH/NIBIB national Biomedical Technology Resource Center supported by P41 EB015897 1S10OD010683 (Johnson). We also acknowledge grant NIH/NIA K01 AG041211 (Badea) and the Duke-Coulter Translational Partnership Grant (Johnson, McNamara). We thank Lucy Upchurch for support of the computing resources.

**Figure Legends**

**Fig. 1. Overview of the VBA processing pipeline.** The VBA pipeline (A) takes multi-modal images, such as MR-DTI contrasts (left inset), and processes them through 8 major stages. The sub-steps for iteratively creating unbiased affine and diffeomorphic targets in Stages 3 and 4 are outlined in (B). Study-specific atlases are generated in addition to statistical maps, while Stage 6 produces regional labels and statistics. Stages 5 and 6 run in parallel, as seen in the unscaled timeline (C). The total runtime is largely determined by the diffeomorphic registrations (Stages 4, 5, and 6), as illustrated by the scaled timeline (D).

**Fig. 2. Overview of the VBA validation framework.** Control and treated images (A) are fed into the VBA pipeline, initialized using our best-guess SyN parameters (B), ultimately producing statistical results (C). During this process, once automated label sets are available for the control images during stage 6 (D), they become input for phantom creation (E). The user can specify how much atrophy or hypertrophy to induce in the structures of their choice. The pipeline is reinitialized, this time with the control and phantom images (F). The results of the phantom VBA (G) are used for calculating several metrics (H), reported alongside the regular VBA results, or used to optimize the SyN or smoothing parameters.

**Fig. 3. Inducing asymmetric morphological changes in control images generates a set of VBM-phantoms.** The label sets (A) of control images generated during Stage 6 were used to create input masks (B) for the left caudate-putamen (CPu, top) and the right hippocampus (Hc, bottom). Localized morphological changes are created by dilating (CPu) and eroding (Hc) the input masks to create target masks (C). The original masks are diffeomorphically registered to the target masks, producing

a warp which relates the original image to the phantom image. The natural log of the Jacobian determinant of the warp (D) reflects the regional volume changes. There is excellent spatial correspondence between the inputs and outputs, with a nominal amount of leakage of the effect size (E) outside the mask regions. The leakage decreases substantially when $p < 0.05$ (F).

**Fig. 4. VBA pipeline runtimes and their relationships with registration parameters and MDT construction strategy.** The workloads of Stages 4 (blue) and 5 (green) for the best-guess (0.25,3,0.5) KA run indicate that a speedup > 7 can be achieved using 6 cluster nodes (A). $Log_{10}$ of the runtimes are plotted for the comparison of the SyN parameters (B). The largest impact (~ 4X) comes from using a SyN parameter of 0.1 instead of 0.25 (B). Also shown are the comparisons for parameters: RegU (C), RegT (D), and MDT group (E).

**Fig. 5. The Dice and phantom metrics reveal the significant impact of RegT.** The left panel of each subplot corresponds to atrophy in the Right Hc, while the right panel characterizes hypertrophy in the Right CPu (Dice coefficients) and Left CPu (phantom metrics). The *x*-axis has been sorted according to the parameter value with the best mean value of that particular metric. Each group or pair with a common location on the *x*-axis represents pipeline runs featuring identical registration/MDT parameters, except for the varying parameter of interest (denoted by "xx"). The first letter identifies the MDT cohort—"C" for controls and "A" for all subjects—while in parentheses are the registration parameters (SyN, RegU, RegT). For both the Dice coefficients of the kainic acid injected mice (A), and the phantom VBA metrics (B-F), increasing RegT from 0 (green) to 0.5 voxels (purple) produced significant improvements. However the absolute distance from target |$\Delta d$| was an exception(B). By this metric, RegT(0.5) was less likely to recover the induced deformations. The same trend of |$\Delta d$| being an outlier amongst other metrics was observed for other parameter

comparisons as well. The arrows highlight Group 10, *A*(0.1,5,xx), chosen for the KA VBA comparison in Figure 6.

**Fig. 6. The impact of RegT on the kainic acid VBA results, illustrated by corrected *q*-maps.** The parameter group *A*(0.1,5,xx) demonstrated very strong effects when varying the RegT parameter. Atrophy (left) and hypertrophy (right) are mapped for three coronal slices. Both atrophy and hypertrophy feature larger clusters for RegT(0.5). The detected hypertrophy is greatly diminished in the contralateral cortex, caudate putamen, hippocampus, and amygdala.

**Fig. 7. The impact of SyN on the kainic acid VBA results, illustrated by corrected q-maps.** The parameter group *A*(xx,3,0.5) demonstrated notable effects when varying the RegT parameter. The number of significant voxels detected was highest using SyN(0.1). Little difference was found between SyN(0.25) and SyN(0.5) with the exception of a small region of atrophy in the ipsilateral hippocampus and adjacent cortex.

**Fig. 8. Comparison of the kainic acid VBA results for the best, median, and poorest performing parameter groups according to the phantom metrics reveals the wide range of potential VBA results.** From left to right, the KA VBA results for the highest (*A*(0.1,3,0.5)), median (*A*(0.25,5,0.5)), and lowest (*A*(0.5,5,0)) rankings of the phantom metrics. This illustrates the variance within the typical parameter space, thus selecting an appropriate set of parameters is critical.

**Fig 9. Histology of the ipsilateral hippocampus using NeuN and GFAP immunoreactivity revealed that KA injected animals present concurrent pathologies**. The yellow box represents

the CA3 hippocampal area, enlarged in the two lower rows. The arrows show: 1) neurodegeneration in the pyramidal cell layers, in the CA1 and in particular CA3 areas (scale bar, 200 µm); 2) granule cells dispersion; 3) astrogliosis (scale bar, 20 µm). These findings support the VBM differences between KA injected and control animals. Abbreviations: Or-oriens layer, Py-pyramidal layer, LMol-lacunosum moleculare, Rad-radiatum, ipsiHc ipsilateral hippocampus.

**Supplementary Material**

**Table S1. Paired *t*-tests comparing Dice coefficients in the kainic acid group for different values of the 4 processing parameters.** Substantial atrophy occurred in the Right Hc and is considered to be "treated," while the Left CPu experienced minimal volumetric change and functions as a control. For SyN (0.1 > 0.25) and (0.1 > 0.5), mild but significant effect sizes were seen in most cases. For SyN (0.25 > 0.5) effect sizes were 2-3x smaller, but still significant. *\*p*-value < 0.01; **corresponding effect size

**Fig. S1. Subject-wise paired *t*-test comparisons of differential changes in the Dice coefficient for 4 structures (columns) of the Kainic Acid mice, as the 4 key parameters are varied (rows).** The insets of the SyN comparison show only the effect size and *p*-values for the (0.1 > 0.25) tests. Varying RegT had the strongest effect on the Dice coefficients, followed by SyN. No discernable differences were detected between RegU(3) and RegU(5) by the Dice. Notably, using the All MDT group was better for detecting the large atrophy in the Right Hc, without incurring a penalty in the other regions.

**Fig. S2. Varying SyN had a modest effect on the various performance metrics.** Closer inspection of the large Dice effects (A) indicated that RegT(0.5) equalized the performances of SyN(0.1) (green), SyN(0.25) (purple), and SyN(0.5) (red), while the consistent drops in Dice for SyN(0.25) and SyN(0.5) were due to RegT(0). With the exception of |Δ*d*| (B), SyN(0.1) had a positive impact on performance on all phantom metrics (C-F). In most cases, using 0.25 instead of 0.5 voxels made a minimal difference. The arrows point to parameter group *A*(xx,3,0.5), which was chosen for KA VBA comparison in Figure 7. Note that this choice shows large differences between the three SyN values across all the phantom metrics—but not Dice.

**Fig. S3. The phantom metrics revealed slight effects when using RegU(3) over RegU(5).** The traditional Dice coefficients (A) did not detect significant difference in performance between the two RegU values. In contrast, the phantom metrics (B-F) all noted a small, yet significant effect size in favor of using RegU(3). Effect sizes were ~2x smaller than those produced by varying SyN. The arrows indicate the group chosen for Figure S4, $A$(0.1,xx,0), because of its large differences in the hypertrophic AUC and TPR values (right panels of E & F).

**Fig. S4. Varying RegU at $A$(0.1,xx,0) produced slight changes in the kainic acid VBA results.** The phantom metrics predicted small variations due to RegU in the KA VBA results of the parameter group $A$(0.1,xx,0). The extent of the significant voxels are consistent with this, with RegU(3) resulting in slightly larger clusters. This is evident in the atrophy in the periventricular regions, for example. Using RegU(5) greatly diminished the hypertrophy detected in the contralateral corpus callosum and cortex.

**Fig. S5. Only the KA Dice coefficients reported a significant advantage of using All subjects to construct the MDT.** MDT(All) greatly improved the Dice values (A) in the region of large deformations. The phantom metrics (B-F) appeared indifferent to the MDT group, indicating that a phantom with larger synthetic volumetric changes would likely result in better correlations between the phantom metrics and the performance of the real KA data. For KA VBA comparison in Figure S6, parameter group $xx$(0.5,5,0.5) (arrows) was selected to illustrate the effects only the Dice coefficients were able to capture.

**Fig. S6. Kainic acid VBA results for the two MDT groups at xx(0.5,5,0.5) showed that using All subjects had an effect on the kainic acid VBA results not indicated by the phantom metrics.** Substantially more localized hippocampal atrophy was detected when using MDT(All). Unexpectedly,

the largest gains in detection were in regions of hypertrophy contralateral to the injection site in the cortex, caudate putamen, amygdala, and hippocampus. MDT(All) detected ipsilateral hypertrophy near the midline and hippocampus, which otherwise would have been unreported. More atrophic affects were detected in the center of the brain when using MDT(Controls). These differences in the KA VBA results were expected based on the Dice coefficients, but were not indicated by the phantom metrics.

**Fig. S7. Correlations between Dice coefficients and the phantom metrics |Δ*d*| (A), *d'* (B), AUC (C), and TPR @ *p* = 0.05 (D) are visualized in scatter plot form.** While statistically significant ($p <$ 0.05) correlations were observed between the phantom metrics and the Dice values, the relationships differed between regions of atrophy and hypertrophy, confounding any generalized relationship between the two. The large respective values of $R$ = 0.708 and 0.836 for the AUC indicate that it is the leading phantom metric for predicting how the Dice coefficients might perform when they are otherwise unavailable.

# References

Ad-Dab'bagh, Y., Lyttelton, O., Muehlboeck, J., Lepage, C., Einarson, D., Mok, K., Ivanov, O., Vincent, R., Lerch, J., Fombonne, E., 2006. The CIVET image-processing environment: a fully automated comprehensive pipeline for anatomical neuroimaging research. Proceedings of the 12th annual meeting of the organization for human brain mapping. Florence, Italy, p. 2266.

ADNI, accessed 5/30/2017. http://adni.loni.usc.edu/wp-content/uploads/2010/05/ADNI2_GE_22_E_DTI.pdf. http://adni.loni.usc.edu/.

Altar, C.A., Baudry, M., 1990. Systemic injection of kainic acid: gliosis in olfactory and limbic brain regions quantified with [3 H] PK 11195 binding autoradiography. Experimental neurology 109, 333-341.

Ashburner, J., Friston, K.J., 2000. Voxel-based morphometry - The methods. Neuroimage 11, 805-821.

Avants, B.B., Epstein, C.L., Grossman, M., Gee, J.C., 2008. Symmetric diffeomorphic image registration with cross-correlation: evaluating automated labeling of elderly and neurodegenerative brain. Medical image analysis 12, 26-41.

Avants, B.B., Kandel, B.M., Duda, J.T., Cook, P.A., Tustison, N.J., Shrinidhi, K.L., 2015. ANTsR: ANTs in R.

Avants, B.B., Tustison, N., Song, G., 2009. Advanced normalization tools (ANTS). Insight J 2, 1-35.

Avants, B.B., Tustison, N.J., Song, G., Cook, P.A., Klein, A., Gee, J.C., 2011. A reproducible evaluation of ANTs similarity metric performance in brain image registration. Neuroimage 54, 2033-2044.

Avants, B.B., Tustison, N.J., Stauffer, M., Song, G., Wu, B., Gee, J.C., 2014. The Insight ToolKit image registration framework. Frontiers in neuroinformatics 8, 44.

Avants, B.B., Yushkevich, P., Pluta, J., Minkoff, D., Korczykowski, M., Detre, J., Gee, J.C., 2010. The optimal template effect in hippocampus studies of diseased populations. Neuroimage 49, 2457-2466.

Badea, A., Ali-Sharief, A., Johnson, G., 2007. Morphometric analysis of the C57BL/6J mouse brain. Neuroimage 37, 683-693.

Badea, A., Gewalt, S., Avants, B.B., Cook, J.J., Johnson, G.A., 2012. Quantitative mouse brain phenotyping based on single and multispectral MR protocols. Neuroimage 63, 1633-1645.

Badea, A., Johnson, G.A., Williams, R., 2009. Genetic dissection of the mouse brain using high-field magnetic resonance microscopy. Neuroimage 45, 1067-1079.

Badea, A., Kane, L., Anderson, R.J., Qi, Y., Foster, M., Cofer, G.P., Medvitz, N., Buckley, A.F., Badea, A.K., Wetsel, W.C., Colton, C.A., 2016. The fornix provides multiple biomarkers to characterize circuit disruption in a mouse model of Alzheimer's disease. Neuroimage 142, 498-511.

Becker, M., Guadalupe, T., Franke, B., Hibar, D.P., Renteria, M.E., Stein, J.L., Thompson, P.M., Francks, C., Vernes, S.C., Fisher, S.E., 2016. Early developmental gene enhancers affect subcortical volumes in the adult human brain. Human brain mapping 37, 1788-1800.

Ben-Ari, Y., Tremblay, E., Ottersen, O., 1980. Injections of kainic acid into the amygdaloid complex of the rat: an electrographic, clinical and histological study in relation to the pathology of epilepsy. Neuroscience 5, 515-528.

Biedermann, S., Fuss, J., Zheng, L., Sartorius, A., Falfán-Melgoza, C., Demirakca, T., Gass, P., Ende, G., Weber-Fahr, W., 2012. In vivo voxel based morphometry: detection of increased hippocampal volume and decreased glutamate levels in exercising mice. Neuroimage 61, 1206-1212.

Blokland, G.A., de Zubicaray, G.I., McMahon, K.L., Wright, M.J., 2012. Genetic and environmental influences on neuroimaging phenotypes: a meta-analytical perspective on twin imaging studies. Twin Research and Human Genetics 15, 351-371.

Borg, J., Chereul, E., 2008. Differential MRI patterns of brain atrophy in double or single transgenic mice for APP and/or SOD. Journal of neuroscience research 86, 3275-3284.

Budin, F., Hoogstoel, M., Reynolds, P., Grauer, M., O'Leary-Moore, S.K., Oguz, I., 2013. Fully automated rodent brain MR image processing pipeline on a Midas server: from acquired images to region-based statistics. Front. Neuroinform 7, 10.3389.

Calabrese, E., Badea, A., Coe, C.L., Lubach, G.R., Shi, Y., Styner, M.A., Johnson, G.A., 2015a. A diffusion tensor MRI atlas of the postmortem rhesus macaque brain. Neuroimage 117, 408-416.

Calabrese, E., Badea, A., Cofer, G., Qi, Y., Johnson, G.A., 2015b. A diffusion MRI tractography connectome of the mouse brain and comparison with neuronal tracer data. Cerebral Cortex 25, bhv121.

Calabrese, E., Badea, A., Watson, C., Johnson, G.A., 2013. A quantitative magnetic resonance histology atlas of postnatal rat brain development with regional estimates of growth and variability. Neuroimage 71, 196-206.

Calabrese, E., Du, F., Garman, R.H., Johnson, G.A., Riccio, C., Tong, L.C., Long, J.B., 2014. Diffusion tensor imaging reveals white matter injury in a rat model of repetitive blast-induced traumatic brain injury. Journal of neurotrauma 31, 938-950.

Camara, O., Schweiger, M., Scahill, R.I., Crum, W.R., Sneller, B.I., Schnabel, J.A., Ridgway, G.R., Cash, D.M., Hill, D.L., Fox, N.C., 2006. Phenomenological model of diffuse global and regional atrophy using finite-element methods. Medical Imaging, IEEE Transactions on 25, 1417-1430.

Chung, M., Worsley, K., Paus, T., Cherif, C., Collins, D., Giedd, J., Rapoport, J., Evans, A., 2001. A unified statistical approach to deformation-based morphometry. Neuroimage 14, 595-606.

Dedeurwaerdere, S., Callaghan, P.D., Pham, T., Rahardjo, G.L., Amhaoul, H., Berghofer, P., Quinlivan, M., Mattner, F., Loc'h, C., Katsifis, A., 2012. PET imaging of brain inflammation during early epileptogenesis in a rat model of temporal lobe epilepsy. EJNMMI research 2, 60.

Dinov, I., Lozev, K., Petrosyan, P., Liu, Z., Eggert, P., Pierce, J., Zamanyan, A., Chakrapani, S., Van Horn, J., Parker, D.S., 2010. Neuroimaging study designs, computational analyses and data provenance using the LONI pipeline. PloS one 5, e13070.

Dinov, I., Van Horn, J., Lozev, K., Magsipoc, R., Petrosyan, P., Liu, Z., MacKenzie-Graha, A., Eggert, P., Parker, D.S., Toga, A.W., 2009. Efficient, distributed and interactive neuroimaging data analysis using the LONI pipeline. Frontiers in neuroinformatics 3, 22.

Dreifuss, S., Vingerhoets, F., Lazeyras, F., Andino, S.G., Spinelli, L., Delavelle, J., Seeck, M., 2001. Volumetric measurements of subcortical nuclei in patients with temporal lobe epilepsy. Neurology 57, 1636-1641.

Eklund, A., Nichols, T.E., Knutsson, H., 2016. Cluster failure: Why fMRI inferences for spatial extent have inflated false-positive rates. Proc Natl Acad Sci U S A 113, 7900-7905.

Ellegood, J., Anagnostou, E., Babineau, B.A., Crawley, J.N., Lin, L., Genestine, M., Dicicco-Bloom, E., Lai, J.K.Y., Foster, J.A., Peñagarikano, O., Geschwind, D.H., Pacey, L.K., Hampson, D.R., Laliberté, C.L., Mills, A.A., Tam, E., Osborne, L.R., Kouser, M., Espinosa-Becerra, F., Xuan, Z., Powell, C.M., Raznahan, A., Robins, D.M., Nakai, N., Nakatani, J., Takumi, T., Van Eede, M.C., Kerr, T.M., Muller, C., Blakely, R.D., Veenstra-Vander Weele, J., Henkelman, R.M., Lerch, J.P., 2015. Clustering autism: Using neuroanatomical differences in 26 mouse models to gain insight into the heterogeneity. Molecular Psychiatry 20, 118-125.

Freeborough, P.A., Fox, N.C., 1997. The boundary shift integral: an accurate and robust measure of cerebral volume changes from registered repeat MRI. Medical Imaging, IEEE Transactions on 16, 623-629.

Friedel, M., van Eede, M.C., Pipitone, J., Chakravarty, M.M., Lerch, J.P., 2014. Pydpiper: a flexible toolkit for constructing novel registration pipelines. Front Neuroinform 8, 67.

Frisoni, G.B., Redolfi, A., Manset, D., Rousseau, M.-É., Toga, A., Evans, A.C., 2011. Virtual imaging laboratories for marker discovery in neurodegenerative diseases. Nature Reviews Neurology 7, 429-438.

Friston, K.J., Holmes, A.P., Worsley, K.J., Poline, J.P., Frith, C.D., Frackowiak, R.S., 1994. Statistical parametric maps in functional imaging: a general linear approach. Human brain mapping 2, 189-210.

Gee, J.C., Reivich, M., Bajcsy, R., 1993. Elastically deforming 3D atlas to match anatomical brain images. Journal of computer assisted tomography 17, 225-236.

Genovese, C.R., Lazar, N.A., Nichols, T., 2002. Thresholding of statistical maps in functional neuroimaging using the false discovery rate. Neuroimage 15, 870-878.

Good, C.D., Ashburner, J., Frackowiak, R.S.J., 2001. Computational neuroanatomy: New perspectives for neuroradiology. Revue Neurologique 157, 797-805.

Gorgolewski, K.J., Auer, T., Calhoun, V.D., Craddock, R.C., Das, S., Duff, E.P., Flandin, G., Ghosh, S.S., Glatard, T., Halchenko, Y.O., Handwerker, D.A., Hanke, M., Keator, D., Li, X., Michael, Z., Maumet, C., Nichols, B.N., Nichols, T.E., Pellman, J., Poline, J.B., Rokem, A., Schaefer, G., Sochat, V., Triplett, W., Turner, J.A., Varoquaux, G., Poldrack, R.A., 2016. The brain imaging data structure, a format for organizing and describing outputs of neuroimaging experiments. Sci Data 3, 160044.

Green, D., Swets, J., 1966. Signal detection theory and psychophysics. 1966. New York 888, 889.

Hayasaka, S., Phan, K.L., Liberzon, I., Worsley, K.J., Nichols, T.E., 2004. Nonstationary cluster-size inference with random field and permutation methods. Neuroimage 22, 676-687.

Hosseini, M.P., Nazem-Zadeh, M.R., Pompili, D., Jafari-Khouzani, K., Elisevich, K., Soltanian-Zadeh, H., 2016. Comparative performance evaluation of automated segmentation methods of hippocampus from magnetic resonance images of temporal lobe epilepsy patients. Medical Physics 43, 538-553.

Jenkinson, M., Beckmann, C.F., Behrens, T.E., Woolrich, M.W., Smith, S.M., 2012. Fsl. Neuroimage 62, 782-790.

Jiang, Y., Johnson, G.A., 2010. Microscopic diffusion tensor imaging of the mouse brain. Neuroimage 50, 465-471.

Johnson, G., Benveniste, H., Black, R., Hedlund, L., Maronpot, R., Smith, B., 1993. Histology by magnetic resonance microscopy. Magnetic resonance quarterly 9, 1-30.

Johnson, G.A., 2000. Three-dimensional morphology by magnetic resonance imaging. Google Patents.

Johnson, G.A., Ali-Sharief, A., Badea, A., Brandenburg, J., Cofer, G., Fubara, B., Gewalt, S., Hedlund, L.W., Upchurch, L., 2007. High-throughput morphologic phenotyping of the mouse brain with magnetic resonance histology. Neuroimage 37, 82-89.

Johnson, G.A., Badea, A., Brandenburg, J., Cofer, G., Fubara, B., Liu, S., Nissanov, J., 2010. Waxholm space: an image-based reference for coordinating mouse brain research. Neuroimage 53, 365-372.

Johnson, G.A., Calabrese, E., Badea, A., Paxinos, G., Watson, C., 2012. A multidimensional magnetic resonance histology atlas of the Wistar rat brain. Neuroimage 62, 1848-1856.

Johnson, G.A., Cofer, G.P., Fubara, B., Gewalt, S.L., Hedlund, L.W., Maronpot, R.R., 2002. Magnetic resonance histology for morphologic phenotyping. Journal of Magnetic Resonance Imaging 16, 423-429.

Jones, D.K., Symms, M.R., Cercignani, M., Howard, R.J., 2005. The effect of filter size on VBM analyses of DT-MRI data. Neuroimage 26, 546-554.

Jovicich, J., Minati, L., Marizzoni, M., Marchitelli, R., Sala-Llonch, R., Bartrés-Faz, D., Arnold, J., Benninghoff, J., Fiedler, U., Roccatagliata, L., Picco, A., Nobili, F., Blin, O., Bombois, S., Lopes, R., Bordet, R., Sein, J., Ranjeva, J.P., Didic, M., Gros-Dagnac, H., Payoux, P., Zoccatelli, G., Alessandrini, F., Beltramello, A., Bargalló, N., Ferretti, A., Caulo, M., Aiello, M., Cavaliere, C., Soricelli, A., Parnetti, L., Tarducci, R., Floridi, P., Tsolaki, M., Constantinidis, M., Drevelegas, A., Rossini, P.M., Marra, C., Schönknecht, P., Hensch, Hoffmann, K.T., Kuijer, J.P., Visser, P.J., Barkhof, F., Frisoni, G.B., 2016. Longitudinal reproducibility of default-mode network connectivity in healthy elderly participants: A multicentric resting-state fMRI study. Neuroimage 124, 442-454.

Karaçali, B., Davatzikos, C., 2006. Simulation of tissue atrophy using a topology preserving transformation model. IEEE transactions on medical imaging 25, 649-652.

Klein, A., Andersson, J., Ardekani, B.A., Ashburner, J., Avants, B., Chiang, M.-C., Christensen, G.E., Collins, D.L., Gee, J., Hellier, P., 2009. Evaluation of 14 nonlinear deformation algorithms applied to human brain MRI registration. Neuroimage 46, 786-802.

Kochunov, P., Lancaster, J.L., Thompson, P., Woods, R., Mazziotta, J., Hardies, J., Fox, P., 2001. Regional spatial normalization: toward an optimal target. Journal of computer assisted tomography 25, 805-816.

Kremen, W.S., Fennema‐Notestine, C., Eyler, L.T., Panizzon, M.S., Chen, C.H., Franz, C.E., Lyons, M.J., Thompson, W.K., Dale, A.M., 2013. Genetics of brain structure: contributions from the Vietnam Era Twin Study of Aging. American Journal of Medical Genetics Part B: Neuropsychiatric Genetics 162, 751-761.

Lee, J., Ehlers, C., Crews, F., Niethammer, M., Budin, F., Paniagua, B., Sulik, K., Johns, J., Styner, M., Oguz, I., 2011. Automatic cortical thickness analysis on rodent brain. SPIE Medical Imaging. International Society for Optics and Photonics, pp. 796248-796248-796211.

Lerch, J.P., Carroll, J.B., Dorr, A., Spring, S., Evans, A.C., Hayden, M.R., Sled, J.G., Henkelman, R.M., 2008. Cortical thickness measured from MRI in the YAC128 mouse model of Huntington's disease. Neuroimage 41, 243-251.

Lerch, J.P., Gazdzinski, L., Germann, J., Sled, J.G., Henkelman, R.M., Nieman, B.J., 2012. Wanted dead or alive? The tradeoff between in-vivo versus ex-vivo MR brain imaging in the mouse. Frontiers in neuroinformatics 6, 6.

Lerch, J.P., Sled, J.G., Henkelman, R.M., 2011. MRI phenotyping of genetically altered mice. Magnetic Resonance Neuroimaging: Methods and Protocols 711, 349-361.

Lévesque, M., Avoli, M., 2013. The kainic acid model of temporal lobe epilepsy. Neuroscience and biobehavioral reviews 37, 2887-2899.

Li, T., Ren, G., Lusardi, T., Wilz, A., Lan, J.Q., Iwasato, T., Itohara, S., Simon, R.P., Boison, D., 2008. Adenosine kinase is a target for the prediction and prevention of epileptogenesis in mice. The Journal of clinical investigation 118, 571-582.

Liu, G., Gu, B., He, X.-P., Joshi, R.B., Wackerle, H.D., Rodriguiz, R.M., Wetsel, W.C., McNamara, J.O., 2013. Transient inhibition of TrkB kinase after status epilepticus prevents development of temporal lobe epilepsy. Neuron 79, 31-38.

Mattes, D., Haynor, D.R., Vesselle, H., Lewellen, T.K., Eubank, W., 2003. PET-CT image registration in the chest using free-form deformations. IEEE transactions on medical imaging 22, 120-128.

Michael, A.M., Evans, E., Moore, G.J., 2016. Influence of Group on Individual Subject Maps in SPM Voxel Based Morphometry. Frontiers in Neuroscience 10, 522.

Minervini, M., Damiano, M., Tucci, V., Bifone, A., Gozzi, A., Tsaftaris, S.A., 2012. Mouse neuroimaging phenotyping in the cloud. Image Processing Theory, Tools and Applications (IPTA), 2012 3rd International Conference on. IEEE, pp. 55-60.

Mouri, G., Jimenez-Mateos, E., Engel, T., Dunleavy, M., Hatazaki, S., Paucard, A., Matsushima, S., Taki, W., Henshall, D.C., 2008. Unilateral hippocampal CA3-predominant damage and short latency epileptogenesis after intra-amygdala microinjection of kainic acid in mice. Brain research 1213, 140-151.

Nichols, T.E., Das, S., Eickhoff, S.B., Evans, A.C., Glatard, T., Hanke, M., Kriegeskorte, N., Milham, M.P., Poldrack, R.A., Poline, J.B., Proal, E., Thirion, B., Van Essen, D.C., White, T., Yeo, B.T.T., 2017. Best practices in data analysis and sharing in neuroimaging using MRI. Nature Neuroscience 20, 299-303.

Nieman, B.J., Bock, N.A., Bishop, J., Chen, X.J., Sled, J.G., Rossant, J., Henkelman, R.M., 2005. Magnetic resonance imaging for detection and analysis of mouse phenotypes. NMR in biomedicine 18, 447-468.

Pagani, M., Damiano, M., Galbusera, A., Tsaftaris, S.A., Gozzi, A., 2016. Semi-automated registration-based anatomical labelling, voxel based morphometry and cortical thickness mapping of the mouse brain. Journal of neuroscience methods 267, 62-73.

Parent, J.M., Timothy, W.Y., Leibowitz, R.T., Geschwind, D.H., Sloviter, R.S., Lowenstein, D.H., 1997. Dentate granule cell neurogenesis is increased by seizures and contributes to aberrant network reorganization in the adult rat hippocampus. Journal of Neuroscience 17, 3727-3738.

Pearson, R., Neal, J., Powell, T., 1986. Hypertrophy of cholinergic neurones of the basal nucleus in the rat following damage of the contralateral nucleus. Brain research 382, 149-152.

R Core Team, 2015. R: A Language and Environment for Statistical Computing. R Foundation for Statistical Computing, Viena, Austria.

Radua, J., Canales-Rodriguez, E.J., Pomarol-Clotet, E., Salvador, R., 2014. Validity of modulation and optimal settings for advanced voxel-based morphometry. Neuroimage 86, 81-90.

Rajagopalan, V., Pioro, E.P., 2015. Disparate voxel based morphometry (VBM) results between SPM and FSL softwares in ALS patients with frontotemporal dementia: which VBM results to consider? BMC neurology 15, 1.

Robbins, S., Evans, A.C., Collins, D.L., Whitesides, S., 2004. Tuning and comparing spatial normalization methods. Medical image analysis 8, 311-323.

Sawiak, S., Wood, N., Williams, G., Morton, A., Carpenter, T., 2009a. SPMMouse: A new toolbox for SPM in the animal brain. ISMRM 17th Scientific Meeting & Exhibition, April, pp. 18-24.

Sawiak, S., Wood, N., Williams, G., Morton, A., Carpenter, T., 2009b. Voxel-based morphometry in the R6/2 transgenic mouse reveals differences between genotypes not seen with manual 2D morphometry. Neurobiology of disease 33, 20-27.

Sawiak, S.J., Wood, N.I., Williams, G.B., Morton, A.J., Carpenter, T.A., 2013. Voxel-based morphometry with templates and validation in a mouse model of Huntington's disease. Magnetic resonance imaging 31, 1522-1531.

Shen, S., Sterr, A., 2013. Is DARTEL-based voxel-based morphometry affected by width of smoothing kernel and group size? A study using simulated atrophy. J Magn Reson Imaging 37, 1468-1475.

Shen, S., Szameitat, A.J., Sterr, A., 2007. VBM lesion detection depends on the normalization template: a study using simulated atrophy. Magnetic resonance imaging 25, 1385-1396.

Smith, S.M., Jenkinson, M., Woolrich, M.W., Beckmann, C.F., Behrens, T.E., Johansen-Berg, H., Bannister, P.R., De Luca, M., Drobnjak, I., Flitney, D.E., 2004. Advances in functional and structural MR image analysis and implementation as FSL. Neuroimage 23, S208-S219.

Thacker, N., 2005. Tutorial: A critical analysis of voxel based morphometry (VBM). Manchester: University of Manchester.

Thompson, P.M., Stein, J.L., Medland, S.E., Hibar, D.P., Vasquez, A.A., Renteria, M.E., Toro, R., Jahanshad, N., Schumann, G., Franke, B., 2014. The ENIGMA Consortium: large-scale collaborative analyses of neuroimaging and genetic data. Brain imaging and behavior 8, 153-182.

Tustison, N.J., 2013. Explicit B-spline regularization in diffeomorphic image registration. Frontiers in neuroinformatics 7, 39.

VanEede, M.C., Scholz, J., Chakravarty, M.M., Henkelman, R.M., Lerch, J.P., 2013. Mapping registration sensitivity in MR mouse brain images. Neuroimage 82, 226-236.

Wang, R., Benner, T., Sorensen, A., Wedeen, V., 2007. Diffusion toolkit: a software package for diffusion imaging data processing and tractography. Proc Intl Soc Mag Reson Med.

Worsley, K.J., Taylor, J., Carbonell, F., Chung, M.K., Duerden, E., Bernhardt, B., Lyttelton, O., Boucher, M., Evans, A.C., 2009. A Matlab toolbox for the statistical analysis of univariate and multivariate surface and volumetric data using linear mixed effects models and random field theory. NeuroImage Organisation for Human Brain Mapping Annual Meeting.

Wuarin, J.-P., Dudek, F.E., 1996. Electrographic seizures and new recurrent excitatory circuits in the dentate gyrus of hippocampal slices from kainate-treated epileptic rats. Journal of Neuroscience 16, 4438-4448.

Yoo, A.B., Jette, M.A., Grondona, M., 2003. Slurm: Simple linux utility for resource management. Job Scheduling Strategies for Parallel Processing. Springer, pp. 44-60.

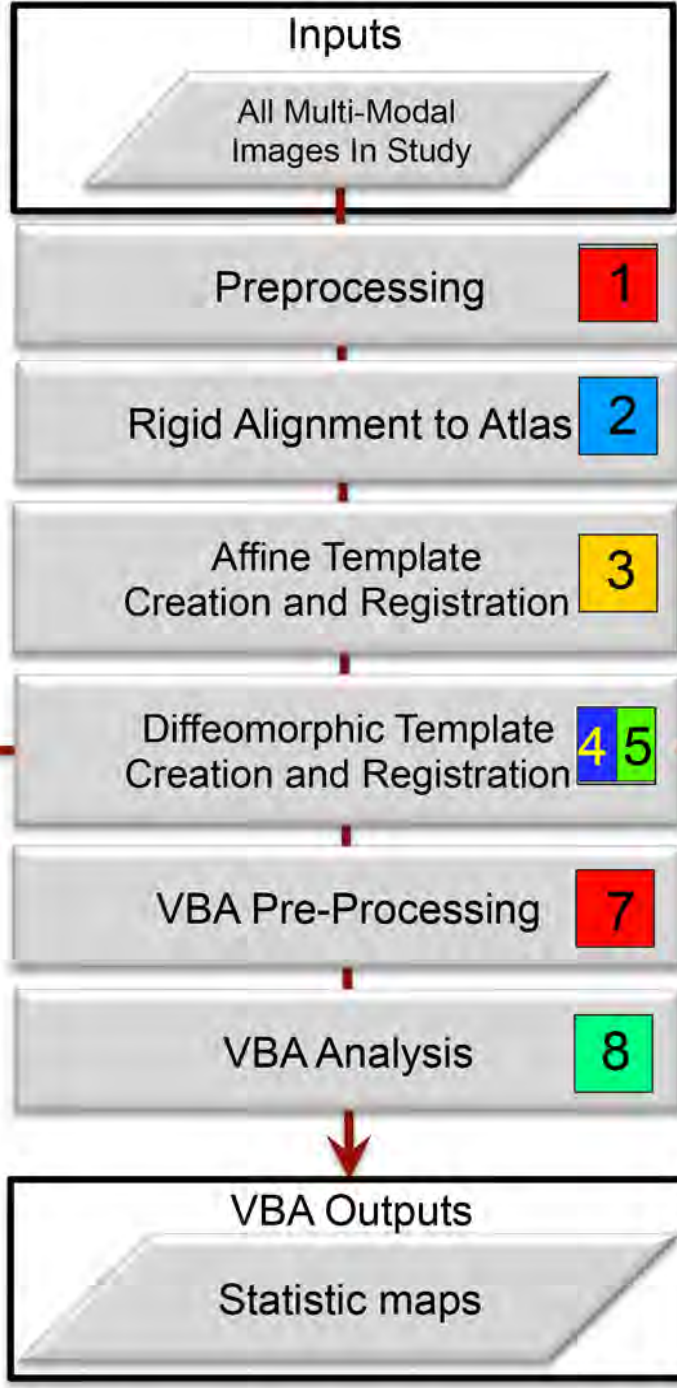
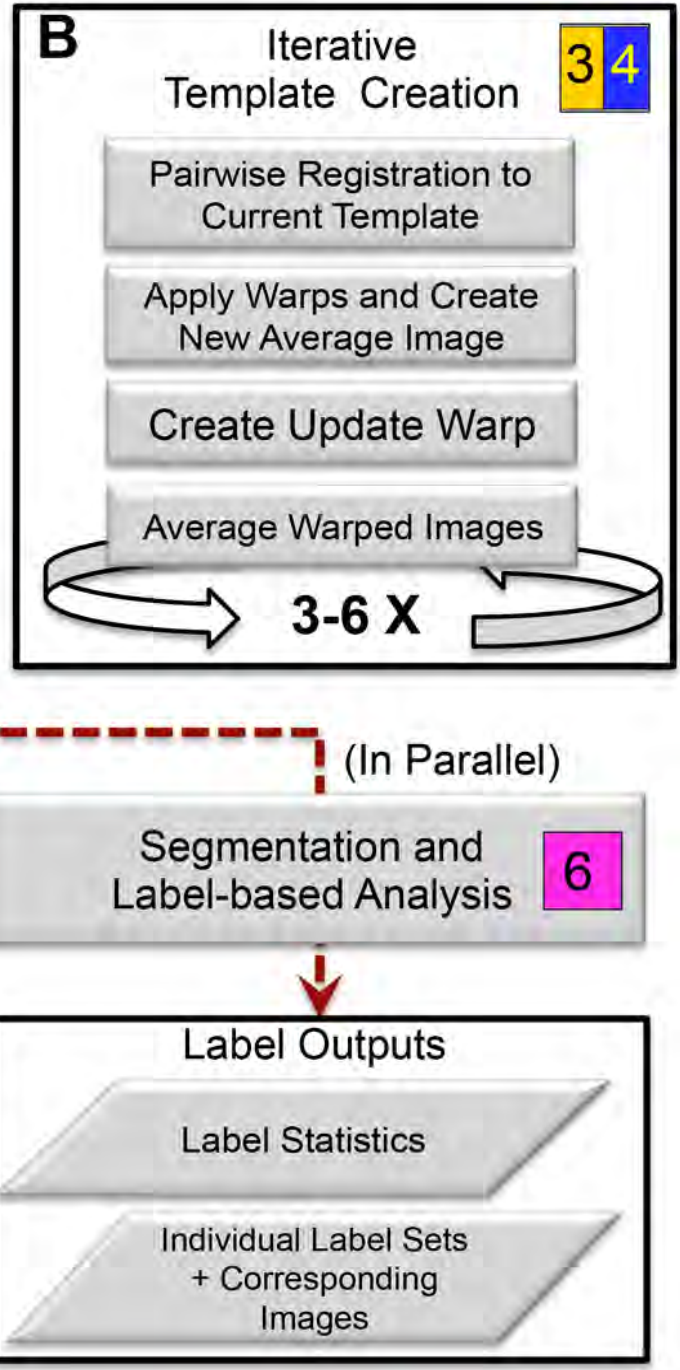
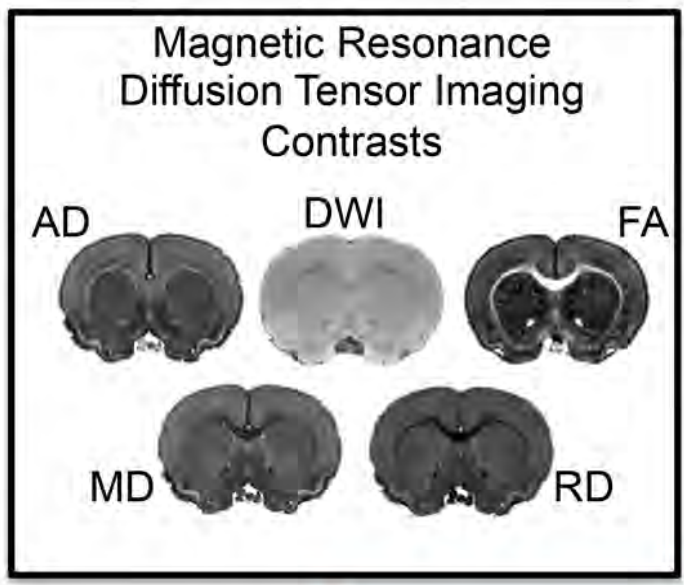
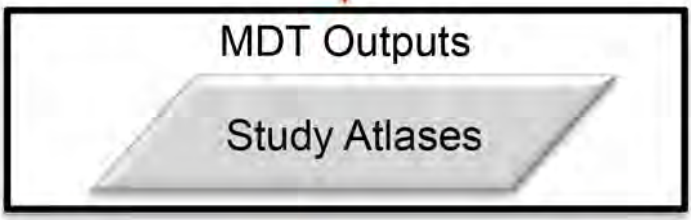
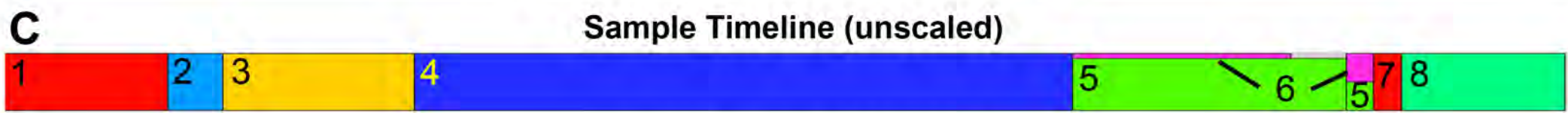
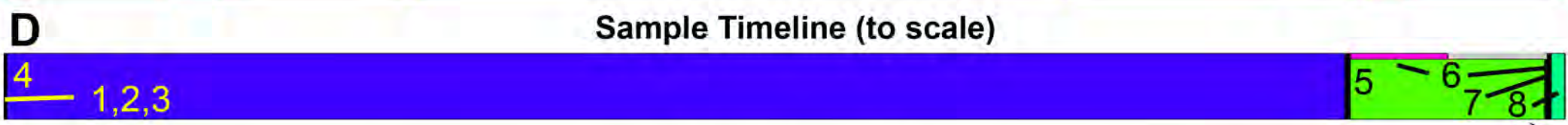

Figure 2

## Phantom Generation

## Phantom VBA

| control labels | input mask | target mask | logJac of warp | effect size | *p*-value |

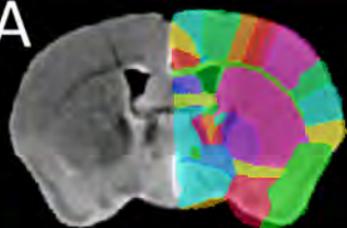 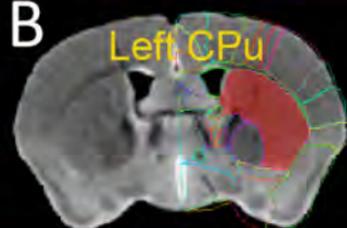 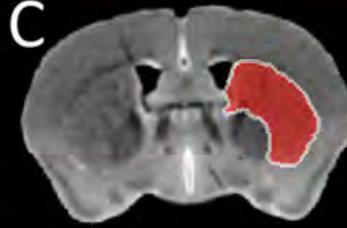 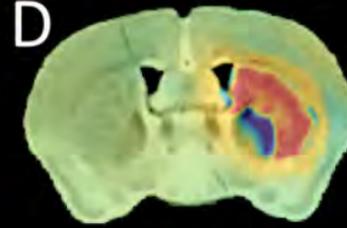 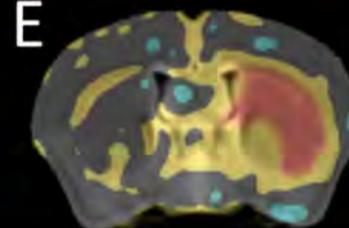 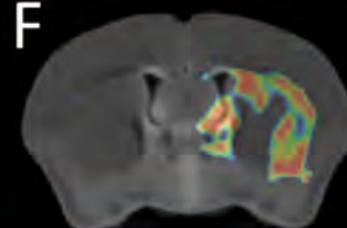

A   B (Left CPu)   C   D   E   F

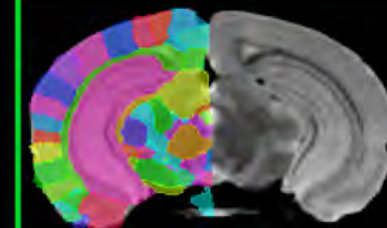 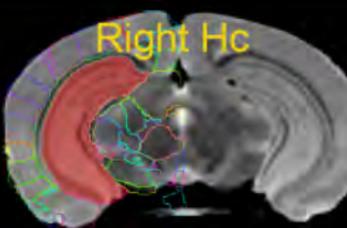 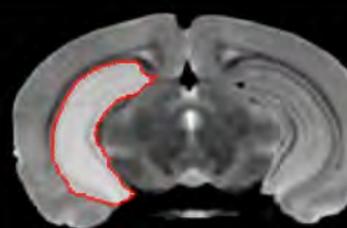 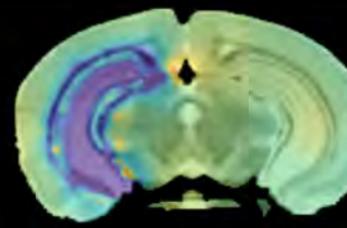 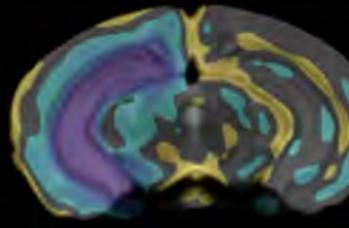 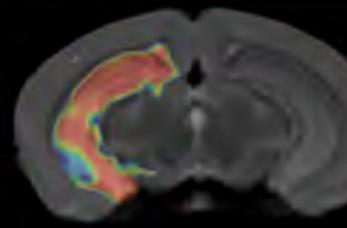

Right Hc

**C:**
Original Mask
Target Mask

**D:**
logJacobian
-0.15 ▬ 0.15

**E:**
effect size
-0.10 ▬ 0.10

**F:**
*p*-value
0.05 ▬ 0

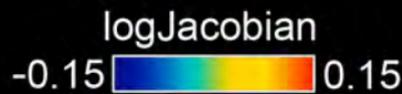 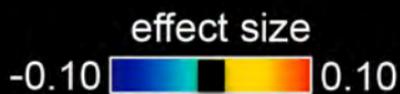 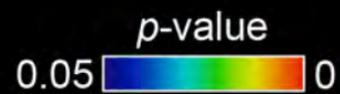 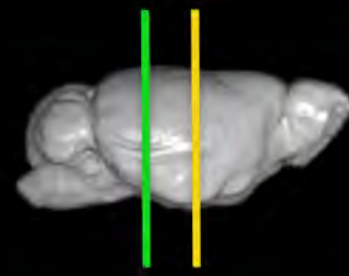 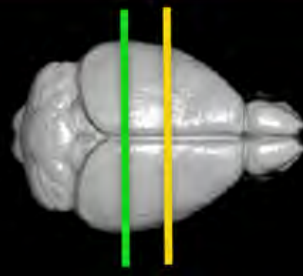

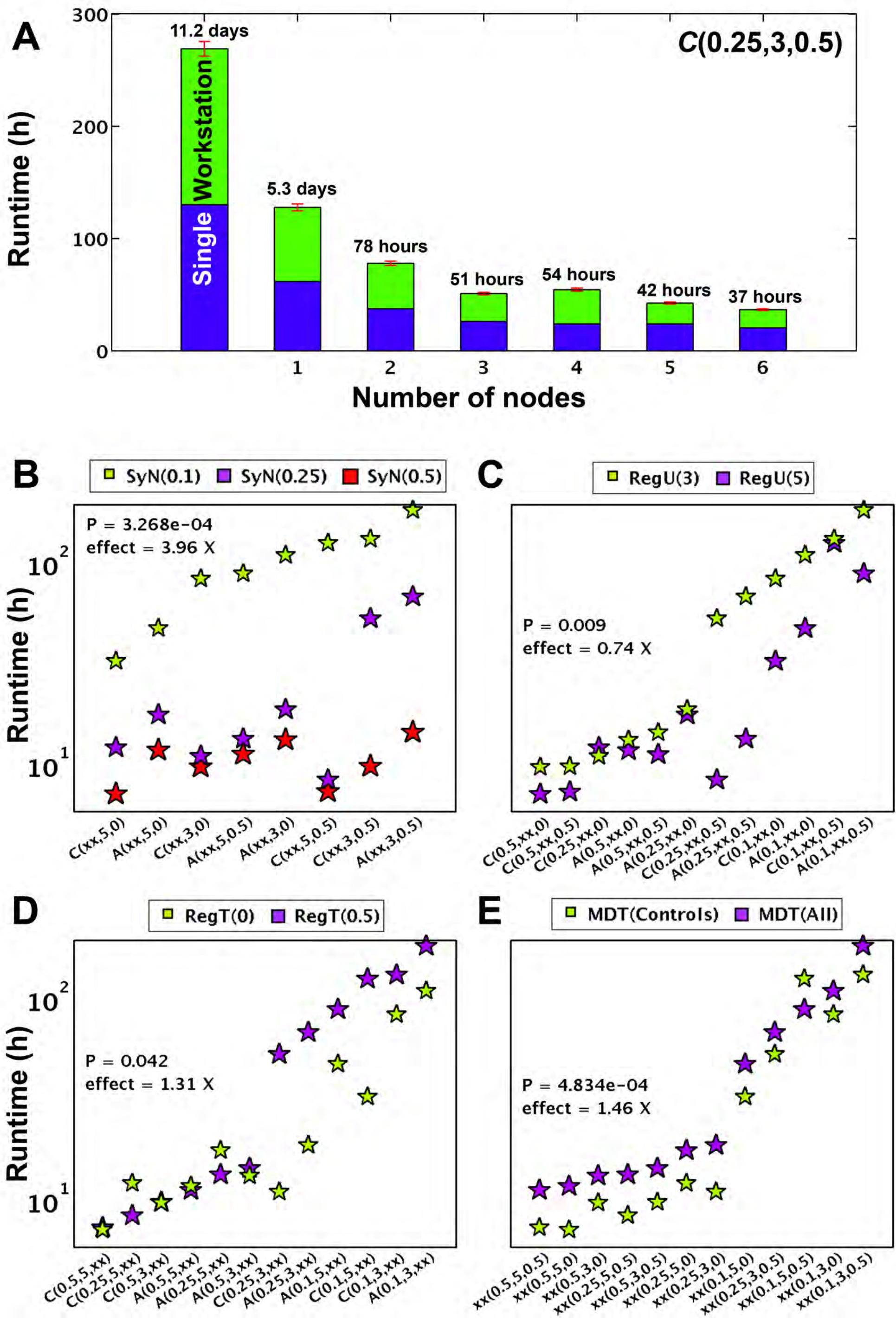

**Figure 5**

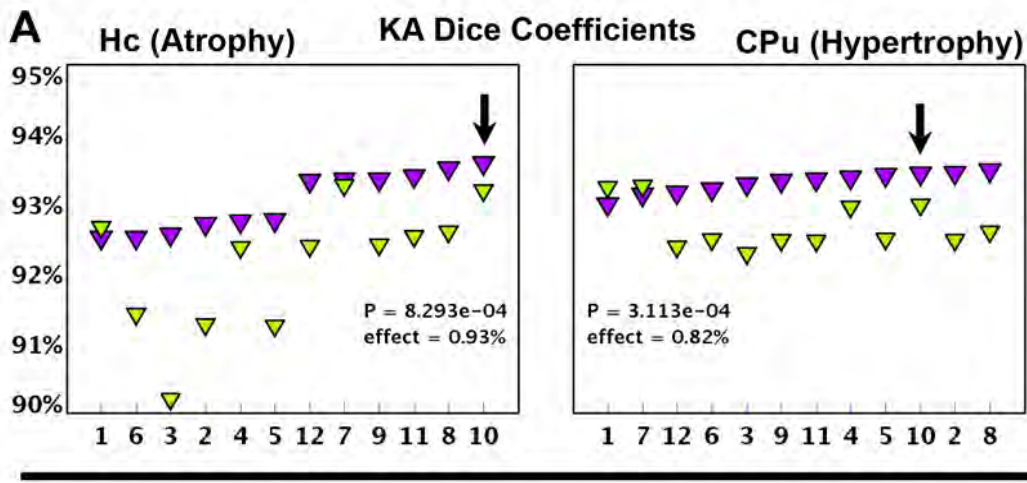

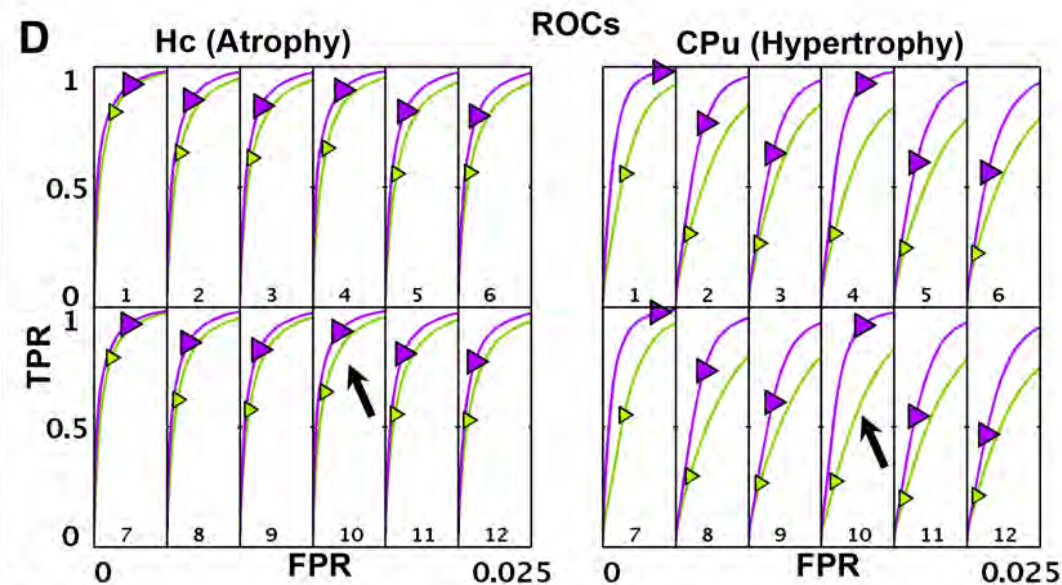

**Legend**  RegT(0)  RegT(0.5)

**A** — Hc (Atrophy) — KA Dice Coefficients — CPu (Hypertrophy)

P = 8.293e-04
effect = 0.93%

P = 3.113e-04
effect = 0.82%

**D** — ROCs — Hc (Atrophy) — CPu (Hypertrophy)

TPR vs FPR

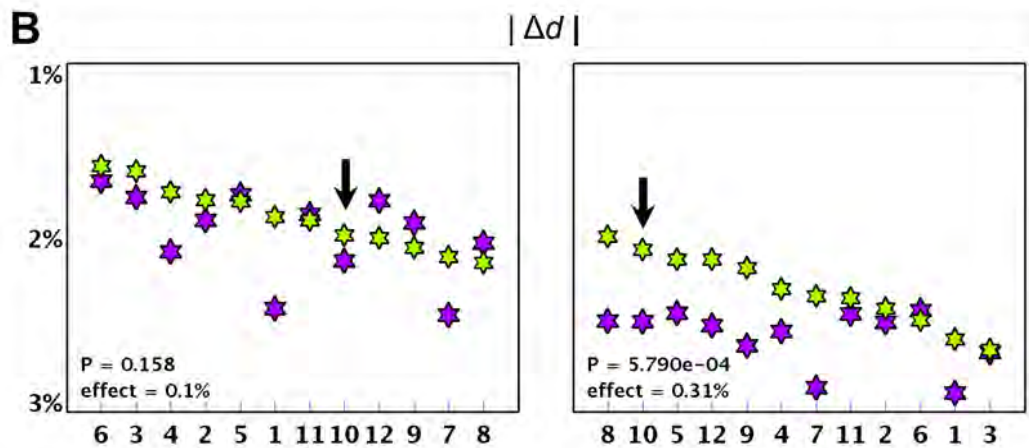

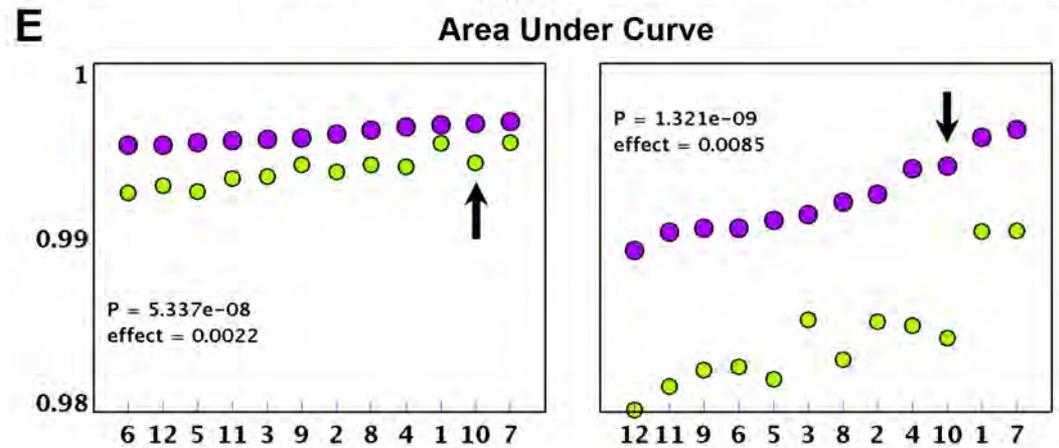

**B** — | Δd |

P = 0.158
effect = 0.1%

P = 5.790e-04
effect = 0.31%

**E** — Area Under Curve

P = 5.337e-08
effect = 0.0022

P = 1.321e-09
effect = 0.0085

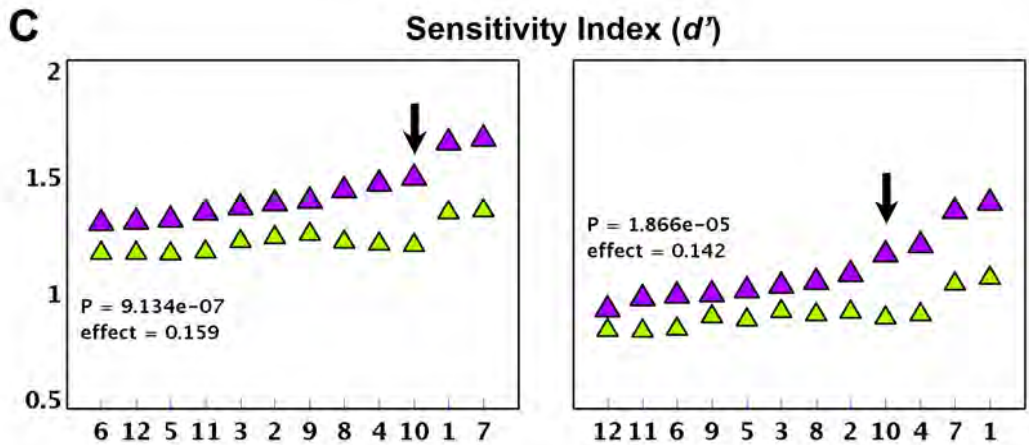

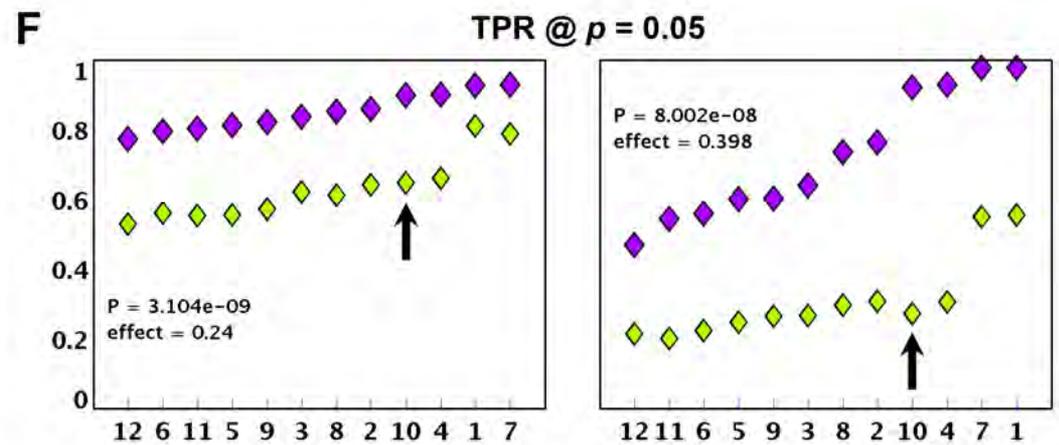

**C** — Sensitivity Index (d′)

P = 9.134e-07
effect = 0.159

P = 1.866e-05
effect = 0.142

**F** — TPR @ p = 0.05

P = 3.104e-09
effect = 0.24

P = 8.002e-08
effect = 0.398

**Parameter Groups:**

1: *C*(0.1,3,xx)   4: *C*(0.1,5,xx)   7: *A*(0.1,3,xx)   10: *A*(0.1,5,xx)

2: *C*(0.25,3,xx)   5: *C*(0.25,5,xx)   8: *A*(0.25,3,xx)   11: *A*(0.25,5,xx)

3: *C*(0.5,3,xx)   6: *C*(0.5,5,xx)   9: *A*(0.5,3,xx)   12: *A*(0.5,5,xx)

Figure 6

# *All*(xx,3,0.5)

**Atrophy**

**Hypertrophy**

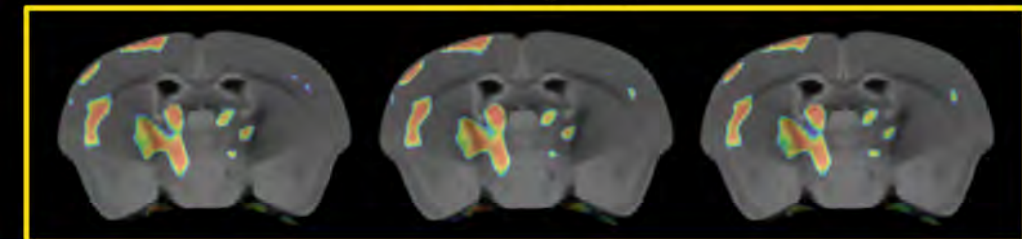 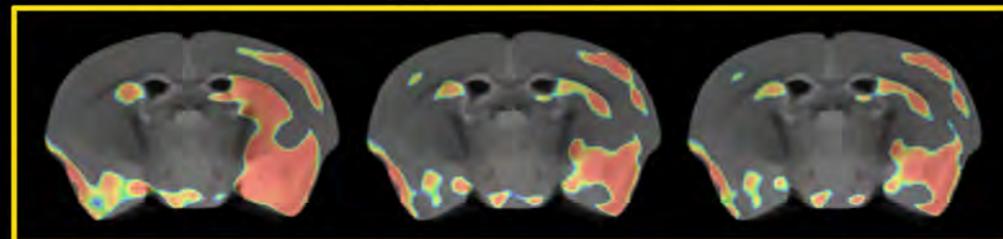

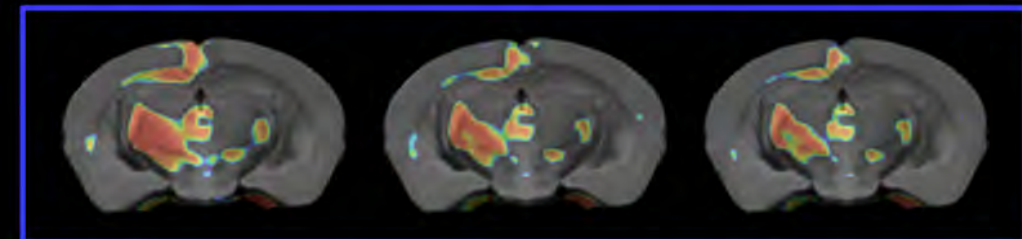 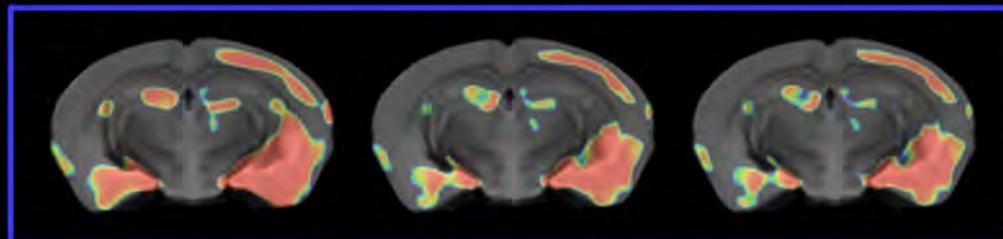

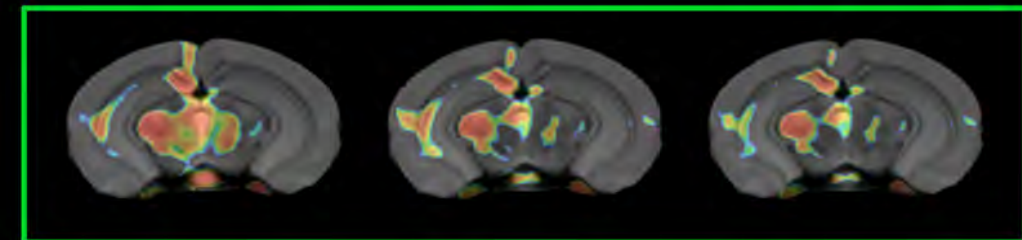 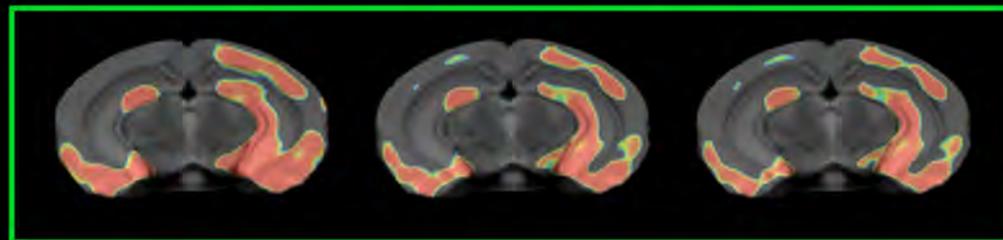

*SyN*(0.1)    *SyN*(0.25)    *SyN*(0.5)          *SyN*(0.1)    *SyN*(0.25)    *SyN*(0.5)

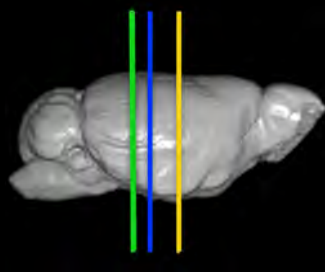 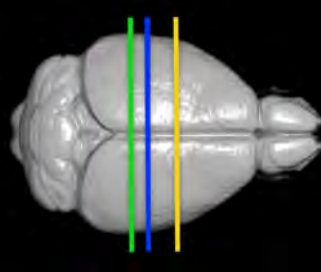 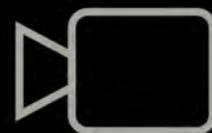 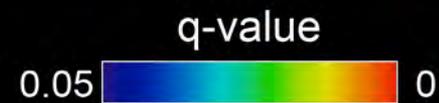

q-value

0.05 ▮ 0

Contrast: log Jacobian

# Atrophy $All(0.1, 5, xx)$ Hypertrophy

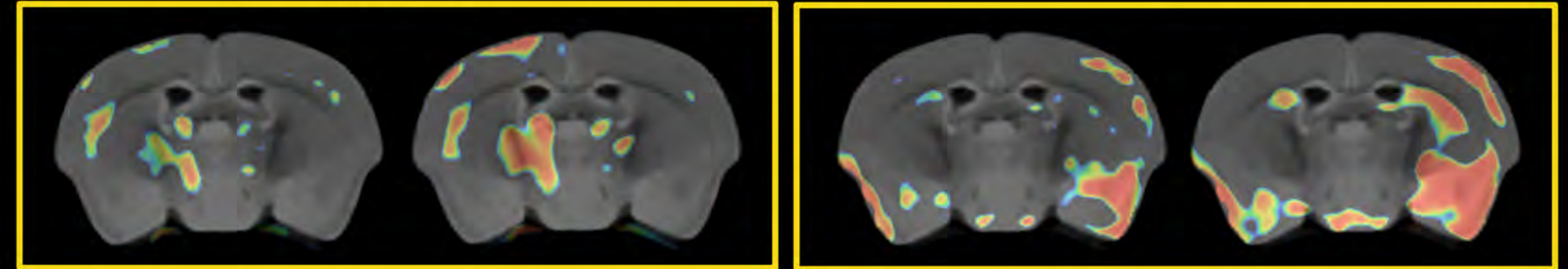

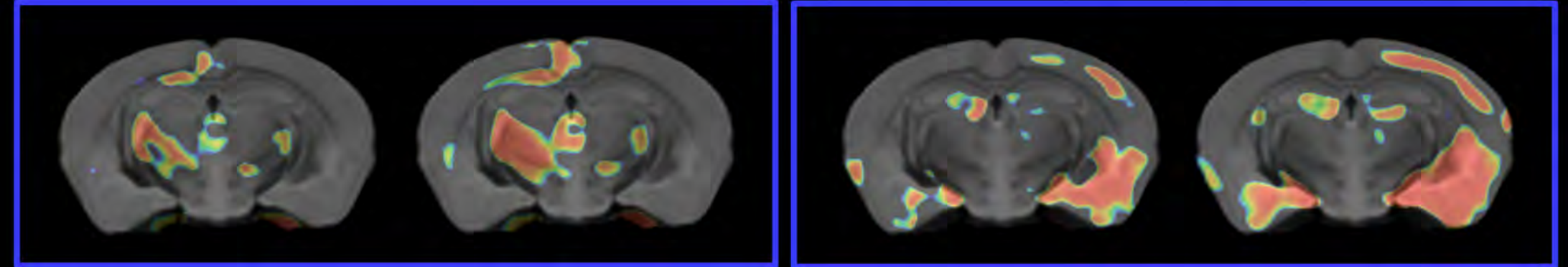

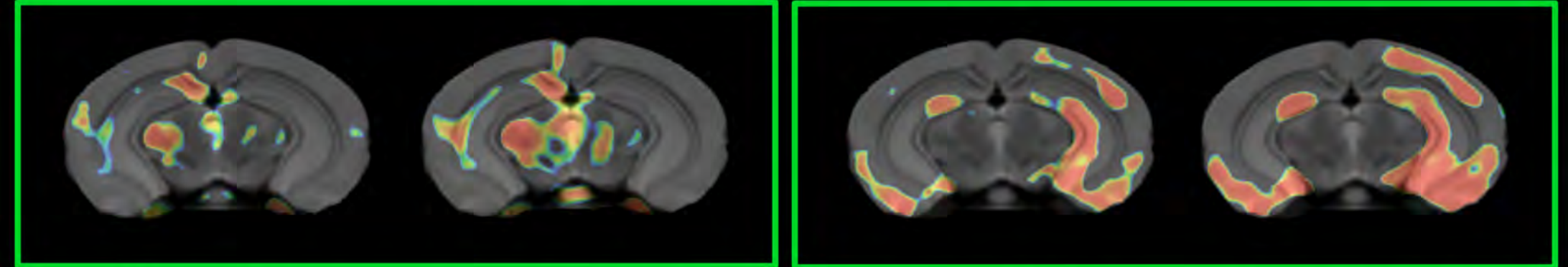

$RegT(0)$  $RegT(0.5)$  $RegT(0)$  $RegT(0.5)$

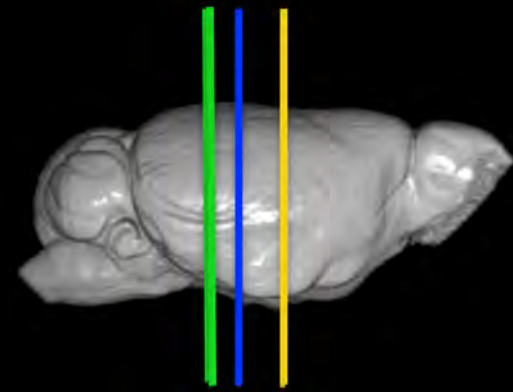
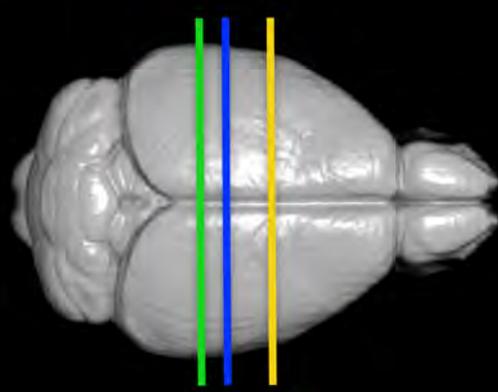
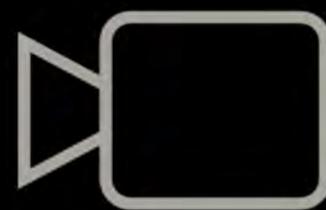

q-value

0.05 [color bar] 0

Contrast: log Jacobian

# Atrophy

# Hypertrophy

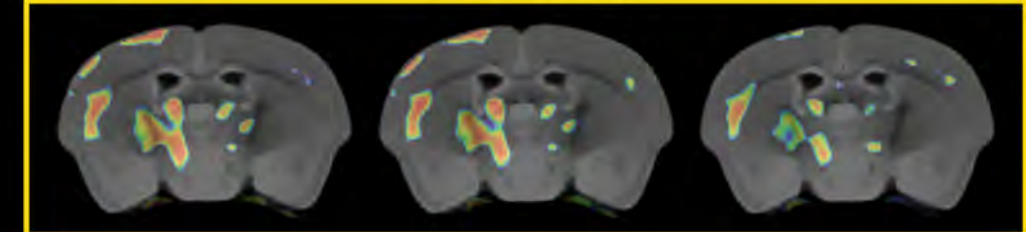 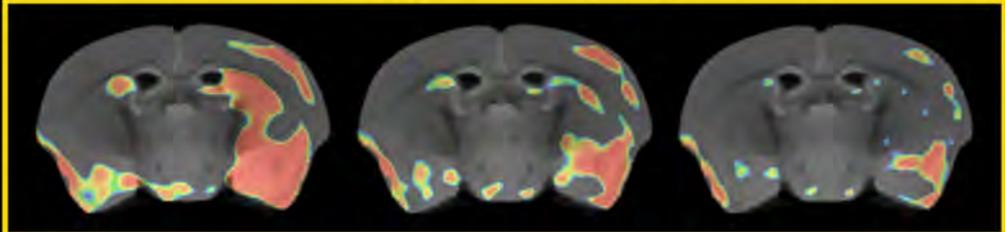

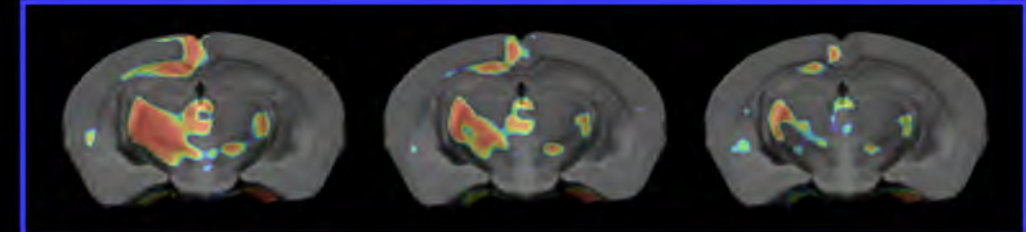 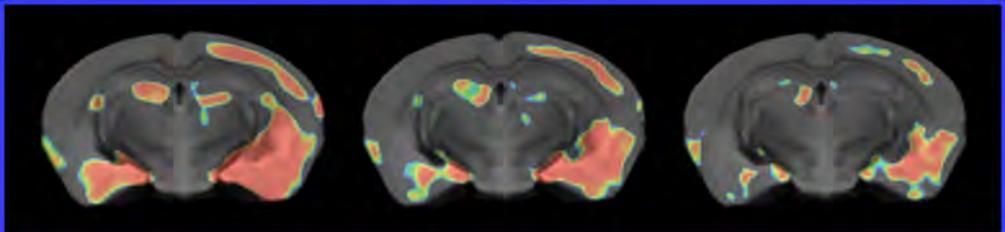

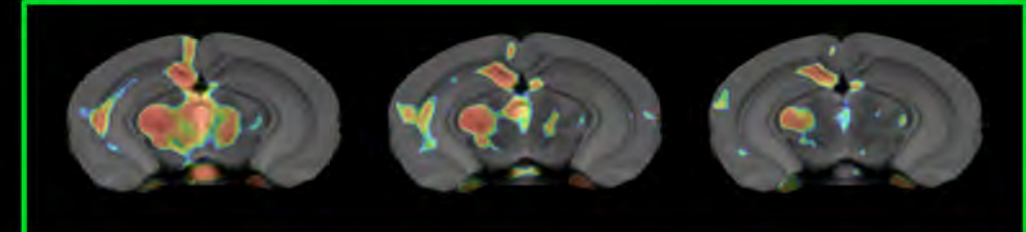 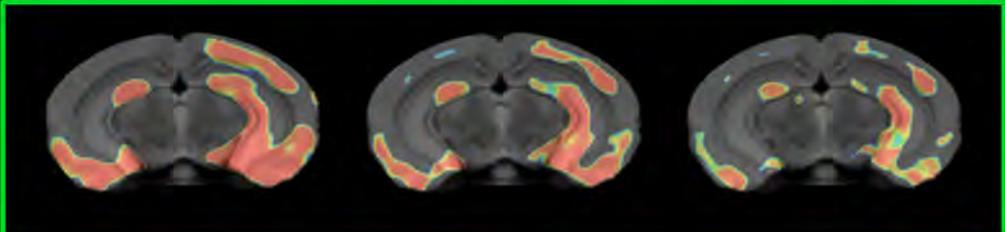

$A(0.1,3,0.5)$    $A(0.25,5,0.5)$    $A(0.5,5,0)$       $A(0.1,3,0.5)$    $A(0.25,5,0.5)$    $A(0.5,5,0)$
#1        #13        #24          #1        #13        #24

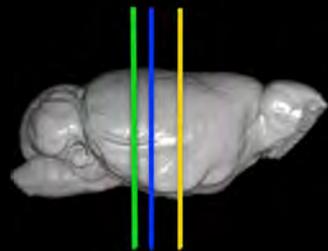 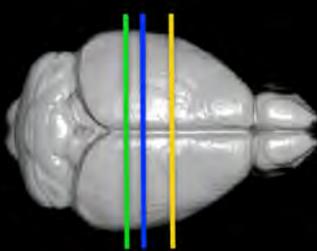 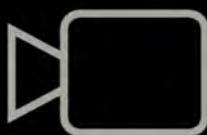

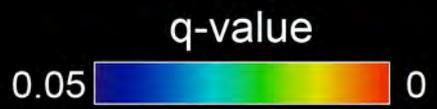

q-value

0.05          0

Contrast: log Jacobian

# PBS

# KA

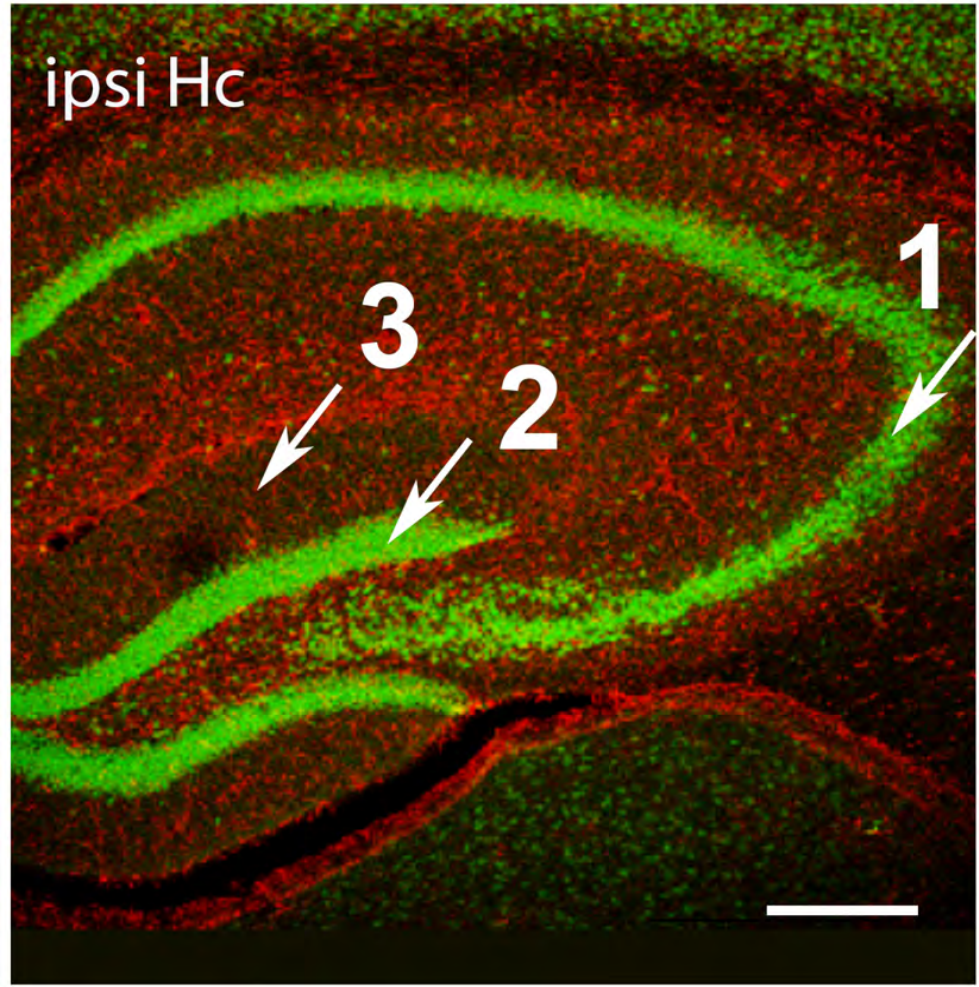

ipsi Hc

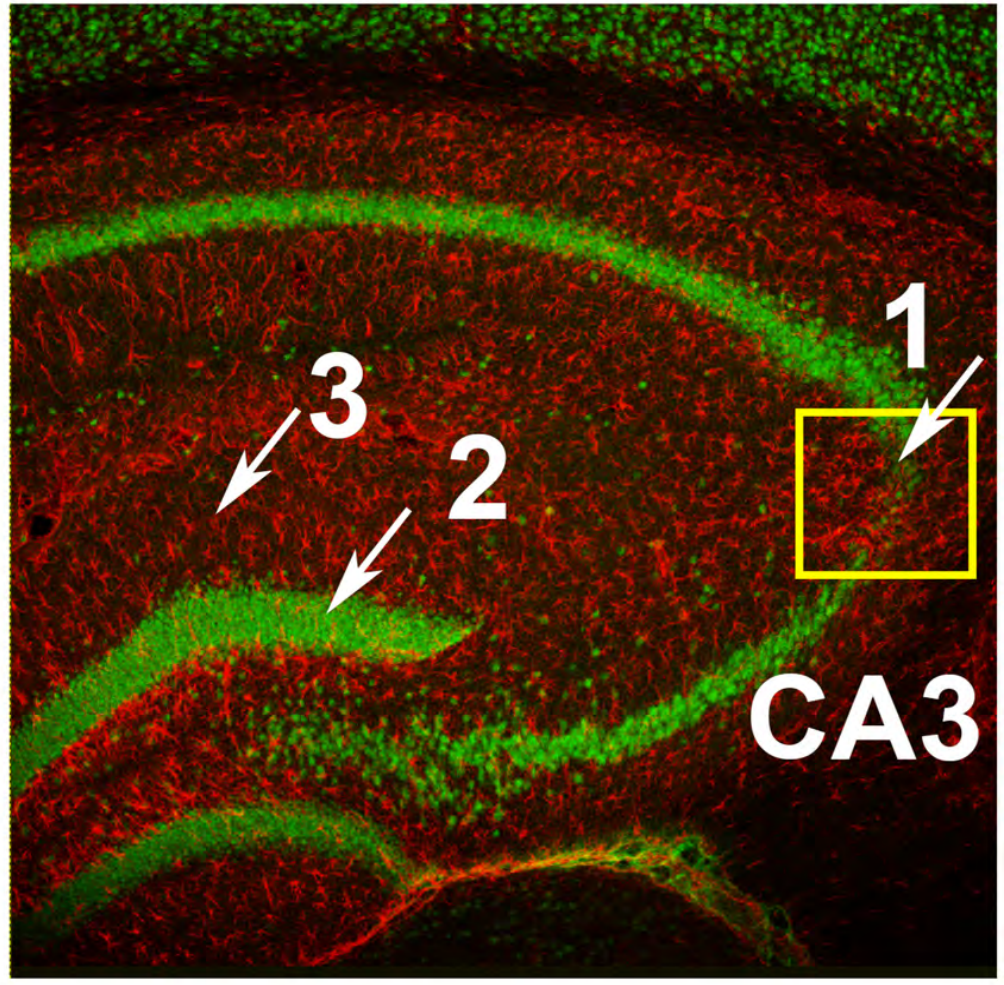

CA3

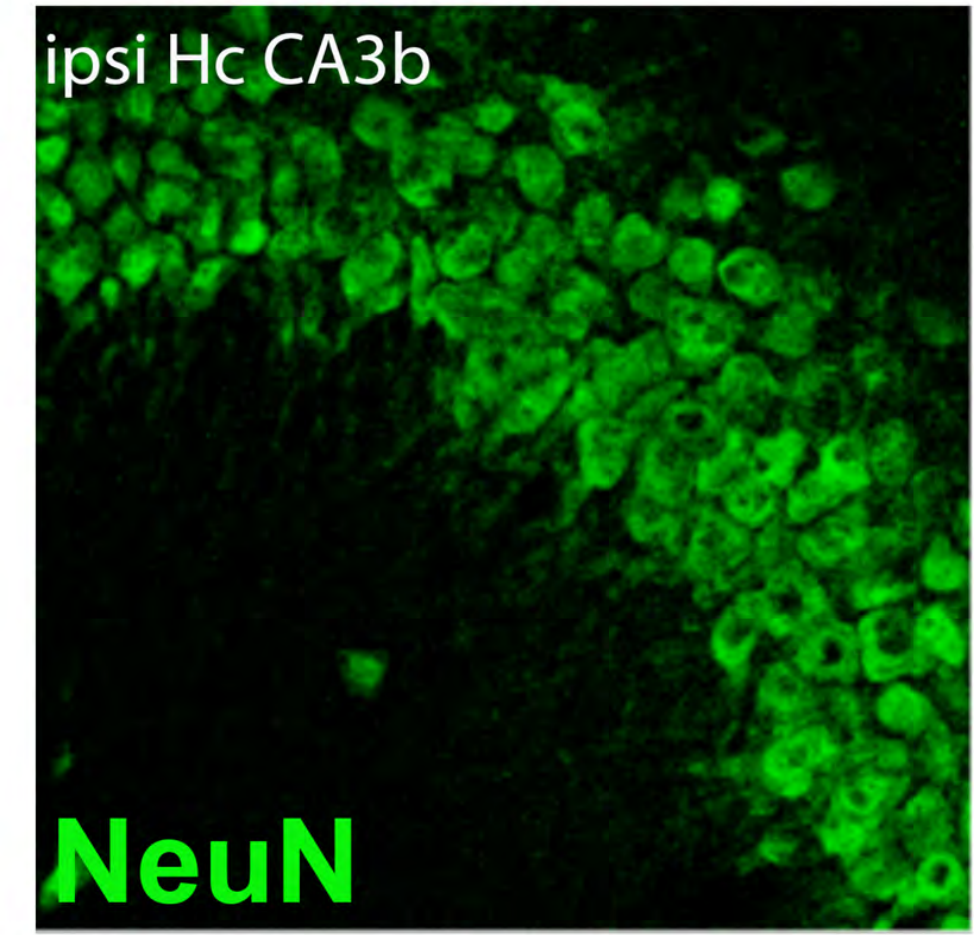

ipsi Hc CA3b

**NeuN**

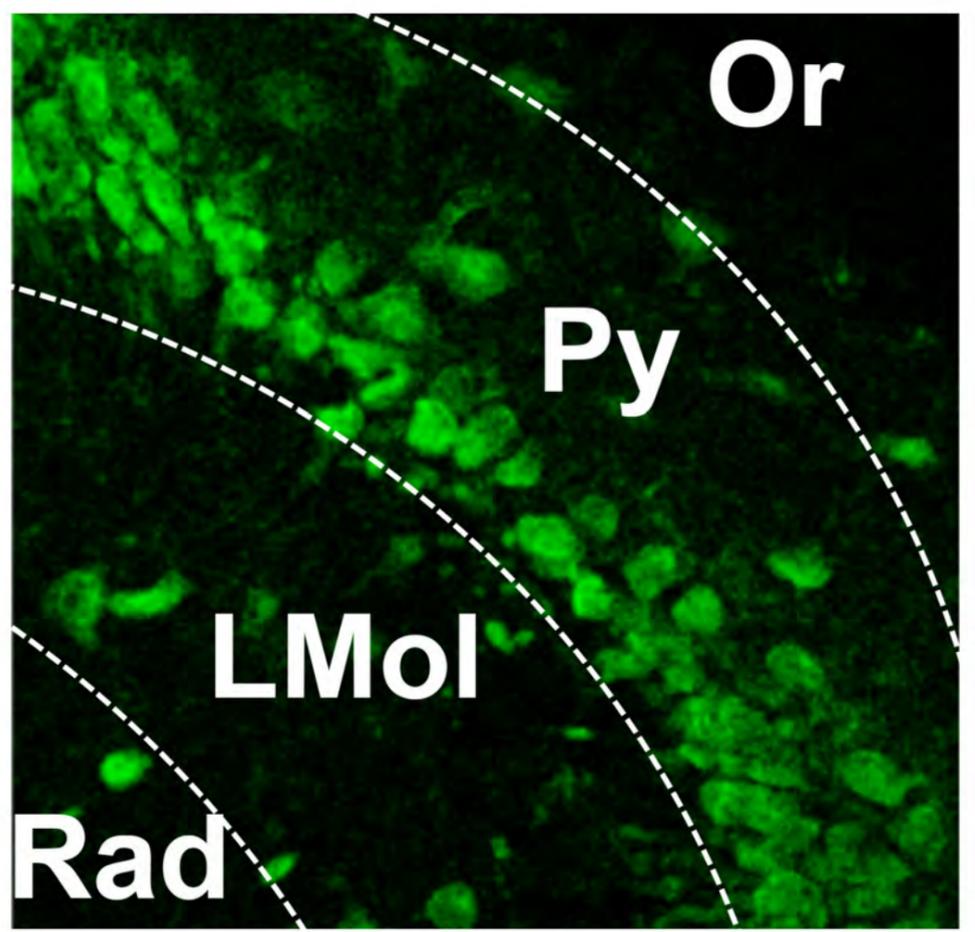

Or

Py

LMol

Rad

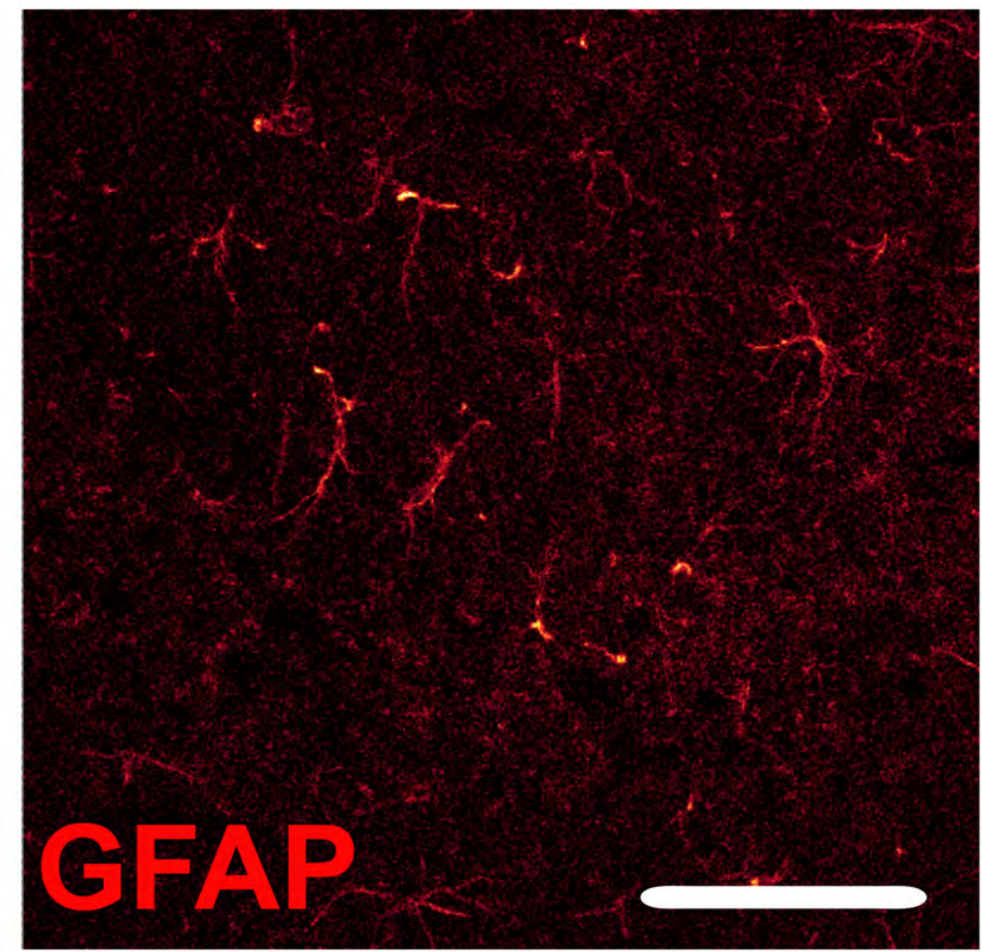

**GFAP**

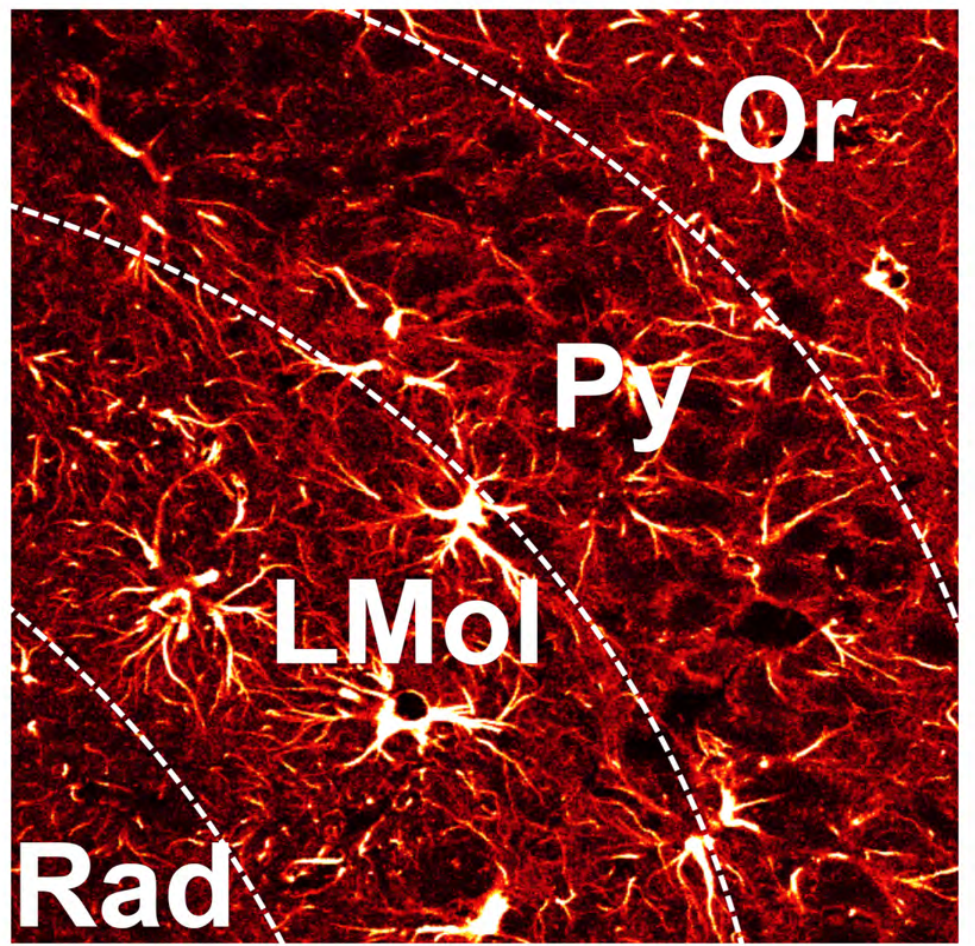

Or

Py

LMol

Rad

| Parameter Comparison | | Left Hc | Right Hc | Left CPu | Right CPu |
|---|---|---|---|---|---|
| **SyN:** | *p*-value: | *5.1e-05 | *9.6e-04 | 0.054 | 0.013 |
| 0.1 > 0.25 | effect size: | **0.32% | **0.40% | 0.30% | 0.24% |
| | *p*-value: | *9.3e-08 | *3.3e-04 | *0.003 | *1.3e-04 |
| 0.1 > 0.5 | effect size: | **0.51% | **0.55% | **0.42% | **0.37% |
| | *p*-value: | *2.2e-10 | 0.023 | *1.5e-07 | *6.0e-07 |
| 0.25 > 0.5 | effect size: | **0.19% | 0.21% | **0.09% | **0.12% |
| **RegU:** | *p*-value: | 0.040 | 0.290 | 0.825 | 0.456 |
| 3 > 5 | effect size: | 0.11% | 0.04% | 0.05% | 0.07% |
| **RegT:** | *p*-value: | *2.1e-14 | *1.2e-08 | *1.6e-11 | *2.4-11 |
| 0.5 > 0 | effect size: | **0.81% | **0.71% | **0.70% | **0.63% |
| **MDT:** | *p*-value: | 0.524 | *1.7e-08 | *8.4e-12 | 0.518 |
| All > Ctrl | effect size: | 0.03% | **0.68% | **0.25% | 0.09% |

**Table S1. Paired *t*-tests comparing Dice coefficients in the kainic acid group for different values of the 4 processing parameters.** Substantial atrophy occurred in the Right Hc and is considered to be "treated," while the Left CPu experienced minimal volumetric change and functions as a control. For SyN (0.1 > 0.25) and (0.1 > 0.5), mild but significant effect sizes were seen in most cases. For SyN (0.25 > 0.5) effect sizes were 2-3x smaller, but still significant. *$p$-value < 0.01; **corresponding effect size

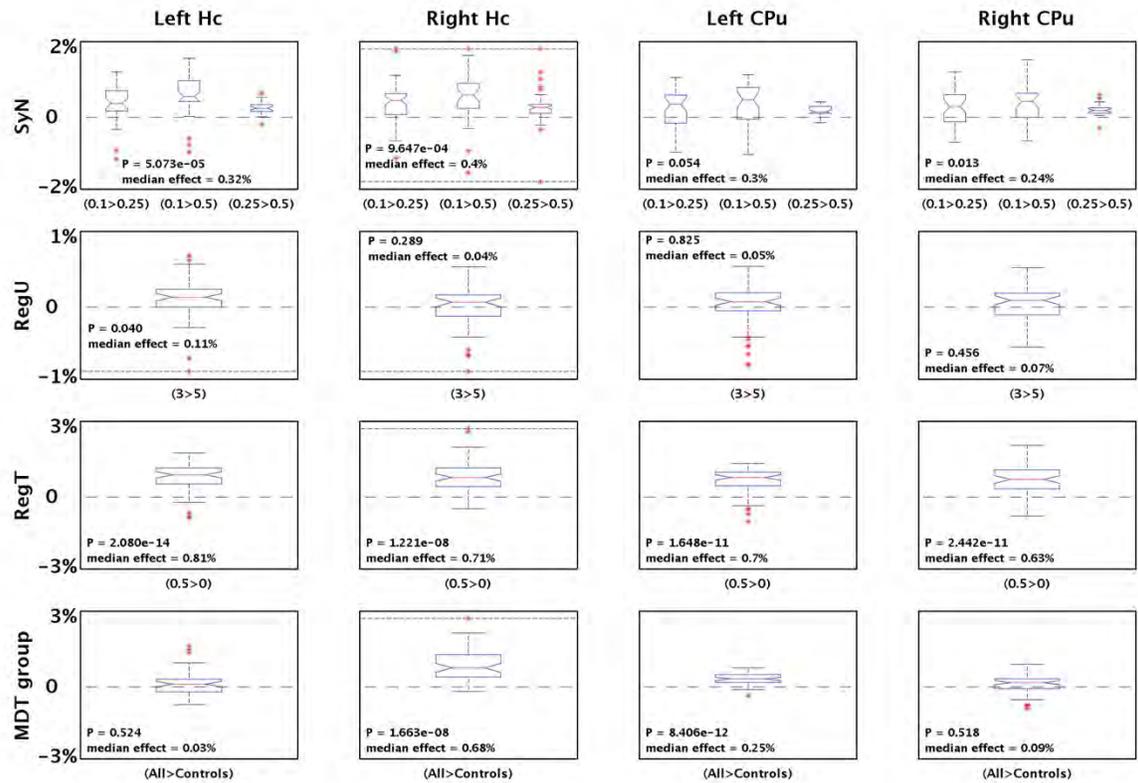

**Fig. S1. Subject-wise paired *t*-test comparisons of differential changes in the Dice coefficient for 4 structures (columns) of the Kainic Acid mice, as the 4 key parameters are varied (rows).** The insets of the SyN comparison show only the effect size and *p*-values for the (0.1 > 0.25) tests. Varying RegT had the strongest effect on the Dice coefficients, followed by SyN. No discernable differences were detected between RegU(3) and RegU(5) by the Dice. Notably, using the All MDT group was better for detecting the large atrophy in the Right Hc, without incurring a penalty in the other regions.

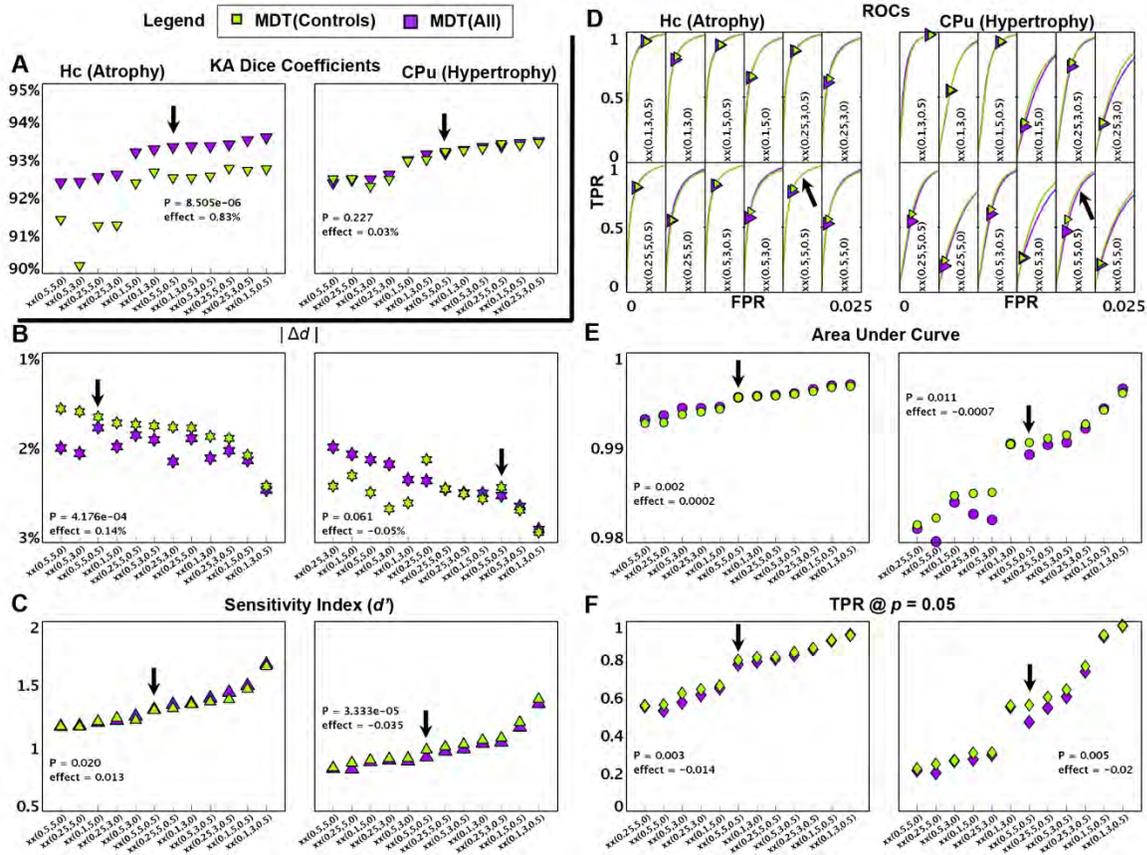

**Fig. S2. Varying SyN had a modest effect on the various performance metrics.** Closer inspection of the large Dice effects (A) indicated that RegT(0.5) equalized the performances of SyN(0.1) (green), SyN(0.25) (purple), and SyN(0.5) (red), while the consistent drops in Dice for SyN(0.25) and SyN(0.5) were due to RegT(0). With the exception of |Δd| (B), SyN(0.1) had a positive impact on performance on all phantom metrics (C-F). In most cases, using 0.25 instead of 0.5 voxels made a minimal difference. The arrows point to parameter group *A*(xx,3,0.5), which was chosen for KA VBA comparison in Figure 7. Note that this choice shows large differences between the three SyN values across all the phantom metrics—but not Dice.

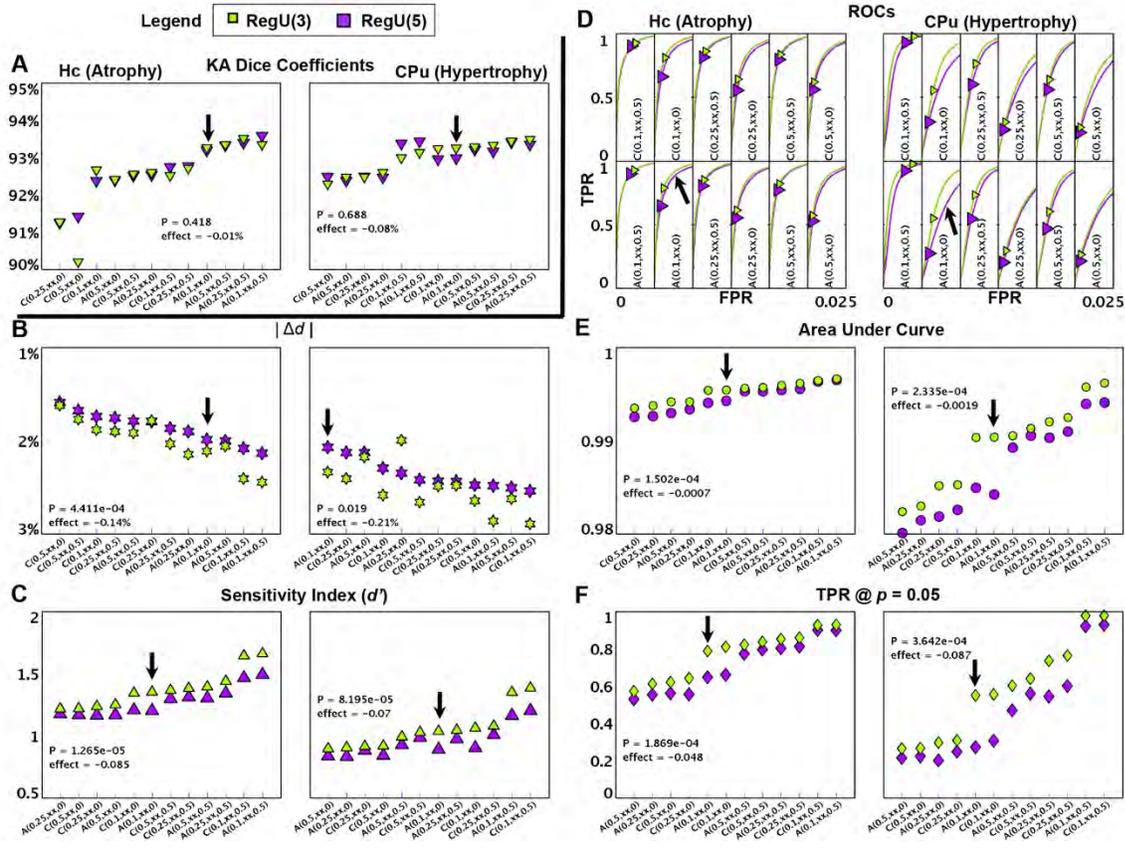

**Fig. S3. The phantom metrics revealed slight effects when using RegU(3) over RegU(5).** The traditional Dice coefficients (A) did not detect significant difference in performance between the two RegU values. In contrast, the phantom metrics (B-F) all noted a small, yet significant effect size in favor of using RegU(3). Effect sizes were ~2x smaller than those produced by varying SyN. The arrows indicate the group chosen for Figure S4, *A*(0.1,xx,0), because of its large differences in the hypertrophic AUC and TPR values (right panels of E & F).

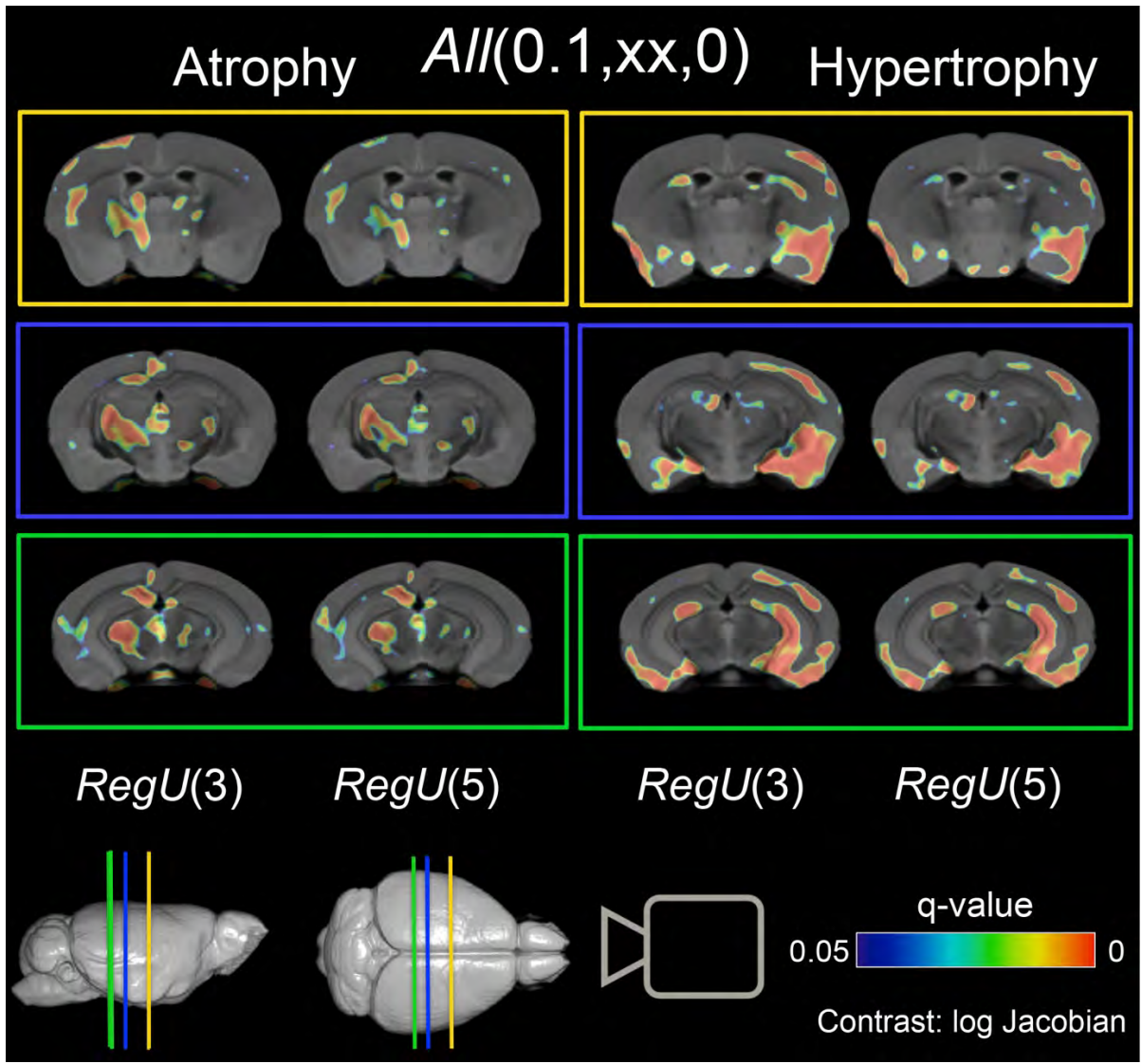

**Fig. S4. Varying RegU at *A*(0.1,xx,0) produced slight changes in the kainic acid VBA results.**
The phantom metrics predicted small variations due to RegU in the KA VBA results of the parameter group *A*(0.1,xx,0). The extent of the significant voxels are consistent with this, with RegU(3) resulting in slightly larger clusters. This is evident in the atrophy in the periventricular regions, for example. Using RegU(5) greatly diminished the hypertrophy detected in the contralateral corpus callosum and cortex.

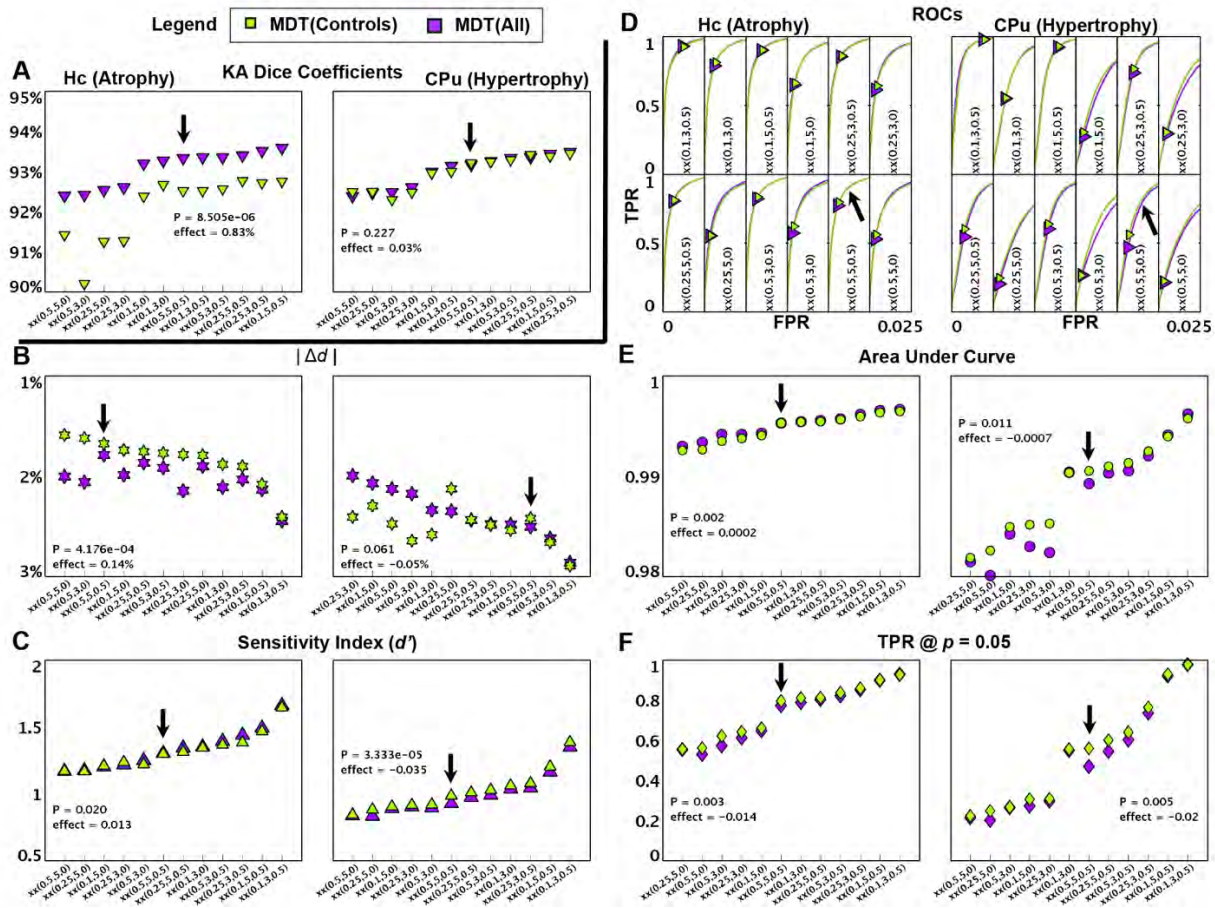

**Fig. S5. Only the KA Dice coefficients reported a significant advantage of using All subjects to construct the MDT.** MDT(All) greatly improved the Dice values (A) in the region of large deformations. The phantom metrics (B-F) appeared indifferent to the MDT group, indicating that a phantom with larger synthetic volumetric changes would likely result in better correlations between the phantom metrics and the performance of the real KA data. For KA VBA comparison in Figure S6, parameter group *xx*(0.5,5,0.5) (arrows) was selected to illustrate the effects only the Dice coefficients were able to capture.

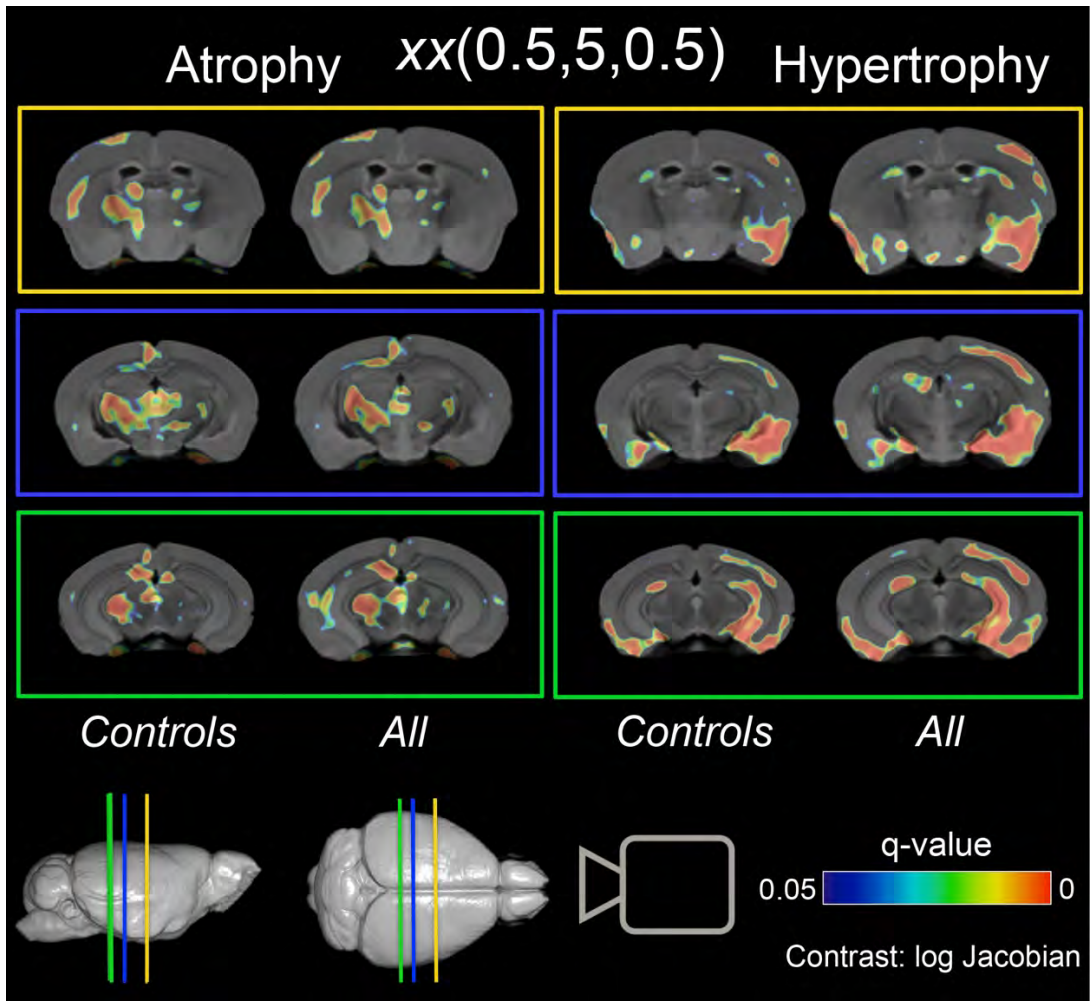

**Fig. S6. Kainic acid VBA results for the two MDT groups at xx(0.5,5,0.5) showed that using All subjects had an effect on the kainic acid VBA results not indicated by the phantom metrics.** Substantially more localized hippocampal atrophy was detected when using MDT(All). Unexpectedly, the largest gains in detection were in regions of hypertrophy contralateral to the injection site in the cortex, caudate putamen, amygdala, and hippocampus. MDT(All) detected ipsilateral hypertrophy near the midline and hippocampus, which otherwise would have been unreported. More atrophic affects were detected in the center of the brain when using MDT(Controls). These differences in the KA VBA results were expected based on the Dice coefficients, but were not indicated by the phantom metrics.

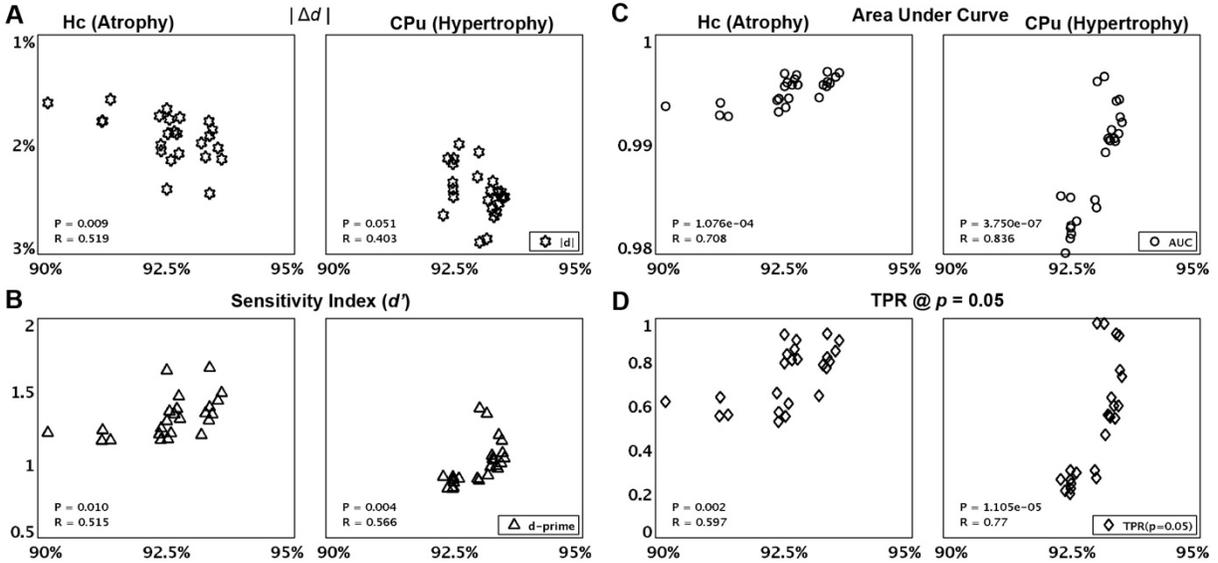

**Fig. S7. Correlations between Dice coefficients and the phantom metrics |Δ*d*| (A), *d'* (B), AUC (C), and TPR @ *p* = 0.05 (D) are visualized in scatter plot form.** While statistically significant (*p* < 0.05) correlations were observed between the phantom metrics and the Dice values, the relationships differed between regions of atrophy and hypertrophy, confounding any generalized relationship between the two. The large respective values of *R* = 0.708 and 0.836 for the AUC indicate that it is the leading phantom metric for predicting how the Dice coefficients might perform when they are otherwise unavailable.